\documentclass[12pt]{article}
\usepackage{epsfig}
\usepackage{graphicx}
\usepackage{cite}
\usepackage{amsfonts}
\usepackage{amssymb}
\usepackage{latexsym}
\usepackage[usenames, dvipsnames]{color}

\def\be{\begin{equation}}
\def\ee{\end{equation}}
\def\ba{\begin{eqnarray}}
\def\ea{\end{eqnarray}}

\def\bq{\begin{quote}}
\def\eq{\end{quote}}

 at 10truept

\newcommand{\beq}{\begin{equation}}
\newcommand{\eeq}{\end{equation}}
\newcommand{\beqa}{\begin{eqnarray}}
\newcommand{\eeqa}{\end{eqnarray}}
\newcommand{\bea}{\begin{eqnarray}}
\newcommand{\eea}{\end{eqnarray}}
\newcommand{\p}{\partial}

\def\lesssim{~\mbox{\raisebox{-.6ex}{$\stackrel{<}{\sim}$}}~}

\def\ltap{\ \raise.3ex\hbox{$<$\kern-.75em\lower1ex\hbox{$\sim$}}\ }
\def\gtap{\ \raise.3ex\hbox{$>$\kern-.75em\lower1ex\hbox{$\sim$}}\ }
\def\gl{\ \raise.5ex\hbox{$>$}\kern-.8em\lower.5ex\hbox{$<$}\ }
\def\roughly#1{\raise.3ex\hbox{$#1$\kern-.75em\lower1ex\hbox{$\sim$}}}

\def\Tr{\textrm{Tr}}
\def\({\left(} \def\){\right)}
\def\[{\left[} \def\]{\right]}

\def\w{\omega}
\def\W{\Omega}

\def\L{\Lambda}

\title{}

\newcommand{\bra}[1]{\langle#1|}
\newcommand{\ket}[1]{|#1\rangle}
\newcommand{\vev}[1]{\langle#1\rangle}

\def\g{\gamma}
\def\half{\frac{1}{2}}
\def\tom{{\tilde \omega}}
\def\tOM{{\tilde \Omega}}

\def\eps{\epsilon}

\renewcommand{\thefootnote}{\fnsymbol{footnote}}

\begin{document}

\thispagestyle{empty}
\begin{flushright}
BRX-TH 6324\\
YITP-17-33\\
September 2017
\end{flushright}
\vspace*{.05cm}
\begin{center}
{\Large \bf Divergences in open quantum systems}

\vspace*{.75cm} {\large Cesar Ag\'on$^{a,b,c}$\footnote{\tt
caagon87@brandeis.edu}  and Albion Lawrence$^{a}$\footnote{\tt
albion@brandeis.edu}}\\
\vspace{.5cm} {\em $^a$Martin Fisher School of Physics, Brandeis University,}\\
{\em Waltham, MA 02453}\\
\vspace{.3cm} {\em $^b$Kavli Institute for Theoretical Physics, University of California,}\\
{\em  Santa Barbara, CA 93106}\\
\vspace{.3cm} {\em $^c$ C.N. Yang Institute for Theoretical Physics, State University of New York,}\\
{\em  Stony Brook, NY 11794-3840}\\

\vspace{.5cm} {\bf Abstract}
\end{center}

We show that for cubic scalar field theories in five and more spacetime dimensions, and for the $T = 0$ limit of the Caldeira-Leggett model, 
the quantum master equation for long-wavelength modes initially unentangled from short-distance modes, and at second order in perturbation theory, contains divergences in the non-Hamiltonian terms.  These divergences ensure that the equations of motion for expectation values of composite operators closes on expectation values of renormalized operators. Along the way we show that initial ``jolt" singularities which occur in the equations of motion for operators linear in the fundamental variables persist for quadratic operators, and are removed if one chooses an initial state projected onto low energies, following the Born-Oppenheimer approximation.


\vfill \setcounter{page}{0} \setcounter{footnote}{0}
\newpage

\tableofcontents

\renewcommand{\thefootnote}{\arabic{footnote}}
\setcounter{equation}{0} \setcounter{footnote}{0}

\section{Introduction}


In classic Wilsonian implementations of coarse-graining in quantum field theory, one diagonalizes the Hamiltonian so that the renormalized coarse-grained variables describe a closed quantum system for low-energy processes.  However, there are a number of cases where this procedure is not appropriate, and the observed long-wavelength degrees of freedom interact with and are entangled with short-distance degrees of freedom, as reviewed in \cite{Agon:2014uxa}.  The observed long-wavelength modes form an open quantum system whose dynamics is governed by a non-Hamiltonian master equation.  This situation occurs in inflationary cosmology \cite{Lombardo:1995fg,Calzetta:1999zr,Burgess:2006jn,Akkelin:2013jsa,Burgess:2014eoa,Burgess:2015ajz}, QCD jet physics \cite{Neill:2015nya}, hydrodynamics
\cite{mori1952quantum,mori1953quantum,zwanzig1960ensemble,zwanzig1961memory,zwanzig2001nonequilibrium,Jeon:1995zm,Jeon:1994if,Calzetta:1998ng,Meyer:1998dg,Minami:2012hs}, and in standard implementations of renormalization in holographic field theories \cite{Agon:2014uxa,Balasubramanian:2012hb}.   

Wilsonian coarse-graining often leads to divergences in the low-energy effective action, due to virtual processes.  This is a standard feature of perturbative quantum field theory, addressed by regularization and renormalization.  The structure of divergences has been less well studied for the case in which the IR and UV are entangled.  For states in which the initial entanglement vanishes, a path integral treatment following Feynman and Vernon \cite{Feynman:1963fq}\ has been applied to interacting scalar quantum field theories in four spacetime dimensions. In this case all of the divergences can be absorbed into the Hamiltonian \cite{Lombardo:1995fg}, up to second order in perturbation theory.  Ref. \cite{Avinash:2017asn}, which appeared as this paper was being completed, studies the renormalization of the Schwinger-Keldysh path integral in a covariant scheme, imposing the condition that the associated master equation be of Kossakowsky-Lindblad type \cite{Kossakowski1972247,Lindblad:1975ef}. In this case the non-Hamiltonian parts of the master equation are in general renormalized.

In this work we study the divergence structure of the quantum master equation using a Hamiltonian framework, for the $T = 0$ Caldeira-Leggett model \cite{Caldeira:1982iu}\ with a "Ohmic" spectrum and for cubic scalar field theory in perturbation theory in arbitrary dimension.  We apply a Hamiltonian regularization closer in spirit to \cite{Kogut:1974ag}, and we assume the fundamental or bare theory is a closed quantum system; the master equation acquires non-Hamiltonian parts due to the coarse-graining, as in \cite{Agon:2014uxa}.  For both systems, we find divergences in the non-Hamiltonian parts of the master equation, and we show that these lead to the equations of motion for expectation values of IR operators closing on renormalized operators. This work is thus complementary to \cite{Lombardo:1995fg,Avinash:2017asn}.

%

In the process of studying the master equations for these systems, we revisit an additional singularity that appears when one takes the standard assumption that the initial state contains no entanglement between the observed ``system" and unobserved ``environment".  In the Caldeira-Leggett model, it is well known that the expectation value of the operators $X,P$ contain a ``jolt" term proportional to $\delta(t)$, where the state is taken to be unentangled at $t = 0$. At the level of these one-point functions, this can be removed without adding any initial entanglement by shifting the state of the environment \cite{RosenaudaCosta:1999uj}.  When one studies the Heisenberg equations of motion for operators quadratic in $X,P$, however, additional ``jolt" terms remain. We show that these largely disappear if one constructs initial states with high-energy modes projected out following the Born-Oppenheimer approximation (to lowest order), in which the environment oscillators are placed in the instantaneous ground state, and argue that they will completely disappear if one performs the projection to higher order.  The aforementioned composite operators divergences remain, however, even after this projection, as we will show in the Caldeira-Leggett example.

\subsection{Overview}

We organize this paper as follows.  In \S2, we study the $T = 0$ Caldeira-Leggett system in considerable depth.
We begin with the exact Heisenberg equations of motion for the observed ``system" variables $X$,$P$, and review the treatment of the initial ``jolt" in \cite{RosenaudaCosta:1999uj}.  We then consider the Heisenberg equations of motion for composite operators $X^2$, $P^2$, and $\{X,P\}$.  In this case the initial jolt survives for initially unentangled states, even after the shift advocated in \cite{RosenaudaCosta:1999uj}. In addition, we find a constant logarithmic divergence in one of the equations, at second order in the system-environment coupling, and we show, using the results in Appendix A, that this acts to ensure the Heisenberg equations of motion closed on renormalized operators.  We then show that if the initial state is constructed following the Born-Oppenheimer approximation to remove (at leading order) high-energy components, the initial jolt is cancelled, though transients remain which we believe are the result of working only to leading order in the Born-Oppenheimer expansion.  

We next study the quantum master equation for this system to second order in perturbation theory, under the assumption of an unentangled initial state, using the formalism developed in \cite{Agon:2014uxa}.  We find that the non-Hamiltonian pieces of the master equation contain time-independent logarithmic divergences. At second order in perturbation theory, the master equation reproduces the Heisenberg equations of motion. The logarithmic divergences are responsible for the time-independent divergences found in the equations of motion for composite operators, and thus act to renormalize those composite operators.


In \S3\ we extend the perturbative treatment of \cite{Agon:2014uxa}\ to scalar quantum field theories in various dimensions, with cubic interactions, with a cutoff on spatial momenta after the fashion of \cite{Kogut:1974ag}.  We find that the non-Hamiltonian part of the master equation contains logarithmic divergences in five spacetime dimensions and linear divergences in six spacetime dimensions (in addition to divergences which renormalize the kinetic term and mass in the Hamiltonian).  The structure of these terms in the master equation is almost identical to the divergent terms in the Caldeira-Leggett model, and we show that they serve the same purpose -- to ensure that the equations of motion for expectation values of composite operators closes on renormalized operators.

We conclude in \S4\ with some discussion of possible future work. 

The Appendices contain calculations that, while important, distract from the narrative flow of the body of the paper. In Appendix \ref{TV}, we compute the singularity structure of the two-point function $\bra{0} X(t_2) X(t_1) \ket{0}$ for the system variable $X$ in the ground state of the Caldeira-Leggett model, and show by taking time derivatives of the above that $\bra{0} P(t_2) P(t_1) \ket{0}$ has a logarithmic divergence as $t_2 \to t_1$. This is subtracted out precisely by the divergences found in \S2,\S3.  In Appendix \ref{F-S} we compute the divergence in $\vev{P^2}$ for a factorized initial state.  In Appendix B we review in more detail the apparent ``jolt" which appears in the Heisenberg equations of motion. In Appendix C, we make contact  with prior work on the Caldeira-Leggett model, by revisiting the master equation when the ``environment" is placed at finite temperature (but still initially unentangled with the observed system), and discuss the subtleties in passing from the standard high-temperature results to the $T = 0$ limit we describe here.

\section{The Caldeira-Leggett model \label{two}}

Our interest is in the structure of the master equation for quantum field theories, and the interpretation of ultraviolet  divergences in that equation.  To set up that interpretation, we begin by considering an exactly solvable quantum mechanical model with a similar divergence structure: the Caldeira-Leggett model \cite{Caldeira:1982iu}\ of a harmonic oscillator called the ``system" linearly coupled to high-frequency  harmonic oscillators called the ``environment".  We study the ``system" as an open quantum system, and compute and interpret the master equation for the density matrix when the combined system and environment degrees of freedom are in a pure state.

The master equation will be computed following \cite{Agon:2014uxa}, for a state with vanishing initial entanglement between the system and environment at time $t = 0$. As we will review below, the master equation is relatively straightforward to compute for such states, and they are of some physical interest.  We will find that the master equation contains divergent terms; one renormalizes the frequency of the system oscillator, while the other is a a logarithmic divergence in the non-Hamiltonian part of the master equation.

The bulk of this section, leading up to this calculation, prepares the ground by showing via more familiar computations that expectation values of system observables in various physical states, including the ground state, carry divergences. When we finally compute the master equation, we can fairly quickly identify and interpret the divergence, and have some confidence that it is not simply an artifact of the initial state.  

In \S2.1\ we will review the Caldeira-Leggett model.  In order to prepare ourself to understand these divergences, we will study the Heisenberg equations of motion for operators that are linear and quadratic in the system coordinate $X$ and momentum $P$.  In \S2.2\ we review the known results for the former,  including the well known ``initial jolt" singularity in these equations, and the understanding and resolution of that singularity in \cite{RosenaudaCosta:1999uj}. 

In \S2.3 we study the Heisenberg equation of motion for quadratic operators. These include ${\cal O} = P^2$, and we will find (using results from Appendix A) that this operator has a divergent expectation value in the true ground state as well as in the factorized initial states we study  -- in fact the divergent piece is identical for both states. 

In \S2.4, we study the expectation value of the Heisenberg equations of motion for the factorized initial state.  After a suitable time averaging, following \cite{Agon:2014uxa}, we find that the terms in the Heisenberg equation arising from mixing between system and environment provide logarithmic singularities that cancel the singularities in the unrenormalized system operators appearing in the Heisenberg equations.  Furthermore, we notice that even if our state satisfies the condition given by \cite{RosenaudaCosta:1999uj}, to which cancel the ``jolt" singularity in the equations for $X$, $P$, a jolt singularity remains, proportional to $\vev{X^2} - \vev{X}^2$.

In \S2.5\ we take a detour and study states in which high-energy components are projected out to lowest order in the Born-Oppenheimer approximation, in which case we expect to see the high-frequency behavior leading to the ``jolt" get cancelled. Indeed, we show that the delta function jolt is cancelled at leading order. As before, the constant logarithmic divergence remains.

Finally, in \S2.6\ we compute the master equation for a class of initial states factorized at $t = 0$.  As stated above, in addition to a divergence in the Hamiltonian part of the equation which renormalizes the system, there is a logarithmic divergence in the non-Hamiltonian part.  In \S2.7, we use the master equation to compute the time evolution of expectation values of quadratic operators, we find that this divergence generates the logarithmic divergence found by direct calculation in \S2.3.  Thus, for the open quantum system, this divergence ensures that the equations of motion for composite operators close on renormalized operators.

\subsection{Review of Caldeira-Leggett Model}

In the Caldeira-Leggett model \cite{Caldeira:1982iu}, the observed ``system" is a particle moving in one dimension with coordinate $X$ and momentum $P$, and the ``environment" variables are some collection of additional particles moving in one dimension with coordinate $x_i$ and momentum $p_i$,  where $i$ labels each individual particle and could be a discrete or continuous variable.
The full Hamiltonian is
\be
\label{sys}
H = \frac{P^2}{2 M}+ \frac{1}{2} M\W^2 X^2
  + \sum_{j=1}^N \left[ \frac{p_j^2}{2m_j} + \frac{1}{2}m_j \omega_j^2 x_j^2 - C_j x_j X\right]
\ee
The dynamics for $X$ depends on the spectrum and state of $x_i$.  We will choose the ``Ohmic" spectrum \cite{Caldeira:1982iu}\ , for which 
\be
	\sum_j \frac{C^2_j}{m_j \omega_j}  f(\w_j)=  \frac{4\gamma M}{\pi}\int_0^{\infty} d\w \, \w F(\w)\ f(\w)\ .
	\label{eq:ohmic}
\ee
Here $F(\w)$ is a cutoff function, removing large frequencies from the spectrum.  The examples we will use are the ``hard" cutoff 
\be
	F(\w) = \theta(\Lambda - \w)\label{eq:hardcut}\ ,
\ee
the ``Drude" cutoff
\be
	F(\w) = \frac{\Lambda^2}{\w^2 + \Lambda^2}\ .\label{eq:drude}
\ee
%

This choice of the spectral density is the same used in \cite{Caldeira:1982iu}, to model quantum Brownian motion. Note that in that work, the environment is held at thermal equilibrium at some temperature $T \gg \Lambda$.  In this work, we are considering the combined system in a pure state, in essence with the environmental variables $T = 0$. (We discuss the relationship between these two limits in Appendix B.)
%
We will find this spectrum leads to divergences in the master equation analogous to that for cubic scalar quantum field theory in four spatial dimensions, due to the growth in the density of oscillators for large $\omega_i$.

In the discussion in \cite{Agon:2014uxa} and the field theory example below, the ``environment" modes are characterized by frequencies high compared to those driving the IR dynamics, that is, $\omega_i \gg \Omega$. This leads to a Born-Oppenheimer-like approximation in which terms in the master equation can be expanded in powers of $\Omega/\omega_i$, after an appropriate time averaging.  Nonetheless, when studying divergences, we will sometimes let the lower limit descend to zero.  This will make certain calculations easier without affecting the ultraviolet divergence structure.

%

\subsection{Heisenberg equations of motion for $X,P$ \label{sec2}}


The equations of motion for $X(t), x_i(t)$ follow simply from (\ref{sys}):
\begin{eqnarray} \label{eq:x}
M \ddot X + M\W^2 X - \sum C_j x_j = 0\label{eq:xsys} \\ \label{eq:xj}
m_j \ddot x_j + m_j \omega_j^2 x_j - C_j X =0
\end{eqnarray}
where $P(t) = M {\dot X}(t)$, $p_i(t) = m_i {\dot x}_i(t)$. To find an equation for the ``system" operator $X$, given initial conditions for the environment, we can solve the second equation (\ref{eq:xj}) by making use of the retarded Green's function for the harmonic oscillator
\be
G_R(t) = \frac{1}{m_j \omega_j} \sin \omega_j t\ \theta(t)\ ,
\ee
to get
\be
x_j(t) = x_j(0) \cos \omega_j t + \frac{p_j(0)}{m_j \omega_j} \sin \omega_j t + \frac{C_j}{m_j \omega_j}\int_0^t ds\ X(s) \sin\(\omega_j (t-s)\) \ .\label{eq:bathsolution}
\ee
Plugging this into (\ref{eq:xsys}) yields
\be \label{eq:x2}
M \ddot X(t) + M\W^2 X(t) - \sum_j C_j x^{0}_j(t) -\sum_j\frac{C^2_j}{m_j \omega_j}\int_0^t ds\ X(s) \sin\(\omega_j (t-s)\) = 0\ .
\ee
where
\begin{eqnarray}
x_j^{0}(t) &=&  x_{j,S} \cos \omega_j t + \frac{p_{j,S}}{m_j \omega_j} \sin \omega_j t \,. \label{eq:F}
\end{eqnarray}
are ${\cal{O}}_S$ are Schr\"odinger picture operators. In (\ref{eq:x2},\ref{eq:F}) we have a linear integro-differential equation with the environment variables acting as an external source, whose effects we can compute from their initial conditions.

To go farther we must pick a spectrum of frequencies $\omega_i$.  For the sake of specificity, we will choose the ``Ohmic" spectrum (\ref{eq:ohmic}). In this case, the last term on the left-hand side of (\ref{eq:x2})\ becomes:
\bea
\label{jolt}
&&\sum_j\frac{C^2_j}{m_j \omega_j}\int_0^t ds\ X(s) \sin\(\omega_j (t-s)\) \nonumber\\
& &\qquad ={4\gamma M}\int_0^t ds\ X(s)\int_0^\infty \frac{d\w}{\pi} \w\sin(\w(t-s)) \nonumber \\
&& \qquad = 4 \gamma M  \int_0^t ds\ X(s)\frac{d \delta(t-s)}{ds}\nonumber\\
& & \qquad = 4 \gamma M\delta(0) X(t) - 4 \gamma M X(0) \delta(t) - 2\gamma M {\dot X}(t)\ ,\label{eq:reshuffle}
\eea
where the last line arises from an integration by parts.  The term $\delta(0)$ in the last line represents the integral 
\be
	\delta(0) =  \int_0^{\infty} \frac{d\omega}{\pi}
\ee
If we cut off the oscillator spectrum at some large frequency $\Lambda$, then we replace $\delta(0) \rightarrow \Lambda/\pi$, so that
\be
	\ddot X(t) +\left(\Omega^2 - \frac{4\gamma\Lambda}{\pi}\right) X(t) + 2\gamma{\dot X}(t) + 4\gamma \delta(t) X(t) = \sum_j \frac{C_j }{M}\,x_j^0(t) \label{eq:syseom}
\ee
The linear divergence thus simply renormalizes the oscillator frequency to 
\be
	\Omega_r^2 = \Omega^2 - \frac{4\gamma\Lambda}{\pi}\
\ee
consistent with \cite{Caldeira:1982iu}: this is the analog of mass renormalization in quantum field theory.   Eq.  (\ref{eq:syseom})\ is an exact operator equation of motion; being linear, it is also the equation of motion for expectation values of $X(t)$ given expectation values of $x_j(0)$, $p_j(0)$. Note the delta function ``jolt" at $t = 0$: we will discuss this in some depth.

If we rewrite (\ref{eq:syseom})\ so that the initial conditions all sit on the RHS, we find:
\be
	\ddot X_H(t) +\Omega_r^2 X_H(t) + 2\gamma{\dot X}_H(t)  = \sum_j\frac{C_j}{M} \, x_j^0(t) - 4 \gamma \delta(t) X_S\label{eq:exactopf}
\ee
where ${\cal{O}}_S = {\cal{O}}(0)$ denotes the Schr\"odinger picture operator. The solution consistent with the initial conditions $X(0) = X_S$, ${\dot X}(0) = P_S/M$ is:
\begin{eqnarray}
	X(t) & = & e^{-\gamma t} \cos({\tilde \Omega} t) X_S \nonumber\\
	& & \qquad \qquad +\ e^{-\gamma t} \sin({\tilde \Omega} t) \left( \frac{P_S}{M{\tilde \Omega}} + \frac{\gamma}{{\tilde\Omega}}\left(3 - 4 \theta(t)\right) X_S\right) \nonumber\\
	& & \qquad \qquad + \int_0^t ds e^{-\gamma(t-s)}\frac{\sin\left({\tilde\Omega}(t-s)\right)}{{M \tilde\Omega}}
		\sum_j C_j x_j^0(s)\label{eq:fullheisx}
\end{eqnarray}
for $t \geq 0$, where $\tilde \Omega^2 = \Omega_r^2 - \gamma^2$. The step function leads to the initial ``jolt" at $t = 0$, where we take $\theta(0) = \half$.

The delta function-supported ``jolt" in (\ref{eq:syseom})\ has been a source of some confusion in the literature. This was (to our mind) partially sorted out in \cite{RosenaudaCosta:1999uj}.  The basic point is that the cross-coupling in (\ref{sys}) for fixed $X(t)$ shifts the minimum of the potential for $x_j(t)$, as we can see if we rewrite (\ref{sys})\ by completing the square for the potential energy of $x_i$:
\be
\label{systwo}
H = \frac{P^2}{2 M}+ \frac{1}{2} M{\W_r}^2 X^2
  + \sum_{j=1}^N \left[ \frac{p_j^2}{2m_j} + \frac{1}{2}m_j \omega_j^2 \left(x_j - \frac{C_j}{m_j \omega_j^2} X\right)^2\right]
\ee

If we allow the initial state for each $x_j$ to be arbitrary -- for example, if the state is centered at $x_j = 0$ when the IR oscillator is centered at some nonzero value of $X$ -- the state generically has components of arbitrarily high energy, leading to the initial kick.  However, since $X$ is a slow-moving mode, we can work in the spirit of the Born-Oppenheimer approximation and choose the initial condition for the ``environmental" modes to minimize the potential for $x_j$ at fixed $X(0)$.
%
%
%
At the level of the one-point functions, if we choose
\begin{eqnarray}
	\vev{x_j(0)} & = & \frac{C_j}{m_j \omega_j^2}\vev{X(0)} \nonumber \\
	\vev{p_j(0)} & = & 0\label{eq:onepointfix}
\end{eqnarray}
then the expectation value of the right hand side of (\ref{eq:exactopf}) cancels. As noted in \cite{RosenaudaCosta:1999uj}, this can be achieved even in the limit that the initial state has vanishing entanglement between the system and environment modes.  

However, once we begin studying the Heisenberg equations for composite operators (which was not done in \cite{RosenaudaCosta:1999uj}), we will find that the above condition does not suffice, and we must take more care to project out rapidly oscillating high-energy modes in order to remove the initial jolt.

%

\subsection{Heisenberg equations for quadratic operators}

The exact equations of motion for quadratic operators close on linear and quadratic operators, and we can therefore make a number of exact statements.  These will introduce new divergences, reflecting the fact that the composite operator $P^2$ requires renormalization.

We consider the following set of IR operators;
\begin{eqnarray}
	O_1 & = & X^2 \nonumber\\
	O_2 & = & XP + PX \nonumber\\
	O_3 & = & P^2
\end{eqnarray}
The equations for $O_1$ are particularly simple:
\be
	\frac{d}{dt} \vev{O_1} = \vev{X{\dot X} + {\dot X X}} = \frac{1}{M}\vev{O_2}
\ee
For $O_2$, we have
\be
	\frac{d}{dt} \vev{O_2} = \frac{2}{M} \vev{O_3} - 2M\Omega^2 \vev{O_1} + \sum_j C_j \vev{X x_j + x_j X}
\ee
We can substitute (\ref{eq:bathsolution}) into the last term, and perform a manipulation along the lines of (\ref{eq:reshuffle})\  to find:
\begin{eqnarray}
	\frac{d}{dt} \vev{O_2} & = & \frac{2}{M} \vev{O_3} - 2 \left(M \Omega_r^2 + 4\gamma M\delta(t)\right) \vev{O_1} - 2\gamma \vev{O_2} \nonumber\\
	& & \qquad \qquad + \sum_j C_j \vev{X x_j^0(t) + x_j^0(t) X}\label{eq:dtxp}
\end{eqnarray}
Note the delta function jolt here.

Finally, a similar calculation yields:
\be
	\frac{d}{dt} \vev{O_3} = - \left(M \Omega_r^2 + 4\gamma M \delta(t)\right)\vev{O_2} - 4\gamma M \vev{O_3} + \sum_j C_j \vev{P x_j^0 + x_j^0 P}\label{eq:dtpp}
\ee
Again, note the delta function jolt term.

%
%

These equations have the form of linear equations for $O_i$, with source terms arising from coupling between the system and environment.  These source terms can be computed explicitly in terms of $X_S$, $P_s$, $x_{i,S}$, $p_{i,S}$ using the operator solutions to the Heisenberg equations in \S2.2, and we will find that they have terms which are divergent as $\Lambda \to \infty$.

To go farther we must study these equations for specific states.  We will focus on two classes; in one, the initial entanglement between $X, x_i$ vanishes; in the second, high-energy modes are projected out up to some level of approximation following the Born-Oppenheimer approximation.  In the first case the delta function jolt is not completely removed, even when the conditions (\ref{eq:onepointfix}) are imposed. In the latter case, the delta function jolt can be removed, although we find some remaining transients that we believe would be removed by working to higher order in the Born-Oppenheimer approximation.  In both cases, there is an additional, logarithmic, time-independent divergence in (\ref{eq:dtxp}), at order $\gamma$.  This divergence, which matches a divergence in the non-Hamiltonian part of the master equation for the factorizes initial state, is a focus of our analysis.

An essential point is that this divergence is {\it not}\ merely a result of a singular initial state. As a roundabout argument, we first note that even in the exact vacuum state, the composite operator $P^2$ has a divergent expectation value and requires renormalization, due to its interaction with the large number of high-frequency modes of the environment. (This divergence is known -- see for example \cite{breuer2007theory}\ and references therein.) In Appendix A we have calculated the short-distance structure of the correlator $\vev{X(t_1) X(t_2)}$ in the true vacuum of the theory.  By taking derivatives, we find a logarithmic divergence in $\vev{P(t_2) P(t_1)}$
\be
	\vev{P(t_2) P(t_1)} = M^2 \p_{t_2} \p_{t_1} \vev{X(t_2) X(t_1)} = - \frac{2\gamma M}{\pi} \ln \Delta t/\tau\label{eq:ppdiv}
\ee
in the limit $\Delta t \to 0$, indicating a divergence in $P^2$ when it is defined by point-splitting.  A further calculation, also in Appendix A, shows that the expectation value of $P^2$ in the factorized initial state we discuss in \S2.4\ has an essentially identical divergence. We will see that this divergence is cancelled precisely by the aforementioned divergence in the final term of (\ref{eq:dtxp}).


\subsection{Factorized initial state}

A standard choice of initial state in the study of open quantum systems is a state with vanishing initial entanglement between the system and environment.  This choice is often made from technical simplicity: there is a particularly straightforward path integral representation \cite{Feynman:1963fq}, and the quantum master equation also takes a 
very specific form, as we will discuss below. However, there are at least two situations in which this is close to the physically correct choice.  The first is the case of an ``interaction quench": for example, at $t < 0$ one sets the couplings between system and environment to zero, and considers the noninteracting ground state as an initial state.   Another situation is that one measures the IR system (or the UV oscillators) completely at time $t = 0$ (that is, measures all of a complete set of commuting observables); in this case, the IR and UV are also disentangled. In practice, of course, the quench or the measurement will take place over some finite time rather than instantaneously, and for many questions it matters whether this quench or measurement happens at scales above or below the cutoff $\Lambda$ (see for example \cite{Das:2014jna,Das:2014hqa,Das:2017sgp}.) We expect the initial jolt will be sensitive to this choice.  We will find, in addition, a constant logarithmic divergence in the master equation and in the Heisenberg equation for system operators: based on study of vacuum divergences in the Appendix A and our lowest-order Born-Oppenheimer treatment below, we believe this divergence is independent of the speed of preparation of the state, and results from the large number of environmental degrees of freedom interacting with the system.

We will consider UV states satisfying (\ref{eq:onepointfix}).  Specifically, defining
\be
	X_0 \equiv {}_{IR}\bra{\psi(0)} {\hat X}_S \ket{\psi(0)}_{IR}\ ,
\ee
we will take the initial wavefunctions for the UV oscillators to be the ground state wavefunction centered on $x_i = \frac{C_i}{m_i\omega_i^2} X_0$:
\be
	\psi(x_i) = \left(\frac{m_i \omega_i}{\pi}\right)^{1/4} e^{- \frac{m_i \omega_i}{2} \left(x_i - \frac{C_i}{m_i \omega_i^2} X_0\right)^2}\ .\label{eq:factinitBO}
\ee

Let us focus on the equation (\ref{eq:dtxp}).  We wish an equation for IR operators alone; given the initial state and the exact solution (\ref{eq:fullheisx})\ for $X(t)$, the final term in (\ref{eq:dtxp}) can be considered as a source term. As above, we consider an Ohmic spectrum of oscillators.  A straightforward calculation yields:
\bea
	\sum_i C_i \langle \{X_H(t),x^0_i(t)\}\rangle & = & 8\gamma M X_0^2 \delta(t) + \sum_{j,k} C_j C_k \int_0^t ds e^{-\gamma(t-s)} \nonumber\\
	& & \qquad \qquad \times \frac{\sin {\tilde \Omega}(t-s)}{\tOM}\langle\{x_j^0(s),x_k^0(t)\}\rangle \nonumber\\
\eea
Plugging this back into equation (\ref{eq:dtxp}), we find:
\begin{eqnarray}
	\frac{d}{dt} \vev{\{X,P\}} & = & \frac{2}{M} \vev{P^2} - 2 M \Omega_r^2\vev{X^2} \nonumber\\
	& & \qquad -  8\gamma M\delta(t) \left(\vev{X^2} - X_0^2\right) - 2\gamma \vev{\{X,P\}} \nonumber\\
	& &  \qquad + \sum_{j,k} C_j C_k \int_0^t ds e^{-\gamma(t-s)}
	\frac{\sin {\tilde \Omega}(t-s)}{\tOM}\langle\{\delta x_j^0(s),\delta x_k^0(t)\}\rangle\nonumber\\
	\label{eq:dtxpfact}
\end{eqnarray}
Note that, it contrast to the equations for the operators $X,P$, the initial condition (\ref{eq:onepointfix})\ does not remove the initial jolt. (And the final term in (\ref{eq:dtxpfact})\ does not depend on the IR state).  Instead, the jolt is proportional to the initial width of the IR wavefunction. The essential point is that the jolt will arise from high-energy components of the state.  To find states without them, we should project out the high-energy states. This can be done systematically starting from the Born-Oppenheimer approximation, as we will discuss in \S2.5\ below.

Using:
\be
	 \langle\{\delta x_j^0(s),\delta x_j^0(t)\}\rangle = \delta_{ij}\frac{\cos\w_j(t-s)}{m_j\w_j}
\ee
and adopting the Ohmic spectrum, the final term in (\ref{eq:dtxpfact}) is:
\bea
	F(t) & = & \sum_{j,k} C_j C_k \int_0^t ds e^{-\gamma(t-s)}
	\frac{\sin {\tilde \Omega}(t-s)}{\tOM}\langle\{\delta x_j^0(s),\delta x_k^0(t)\}\rangle\nonumber\\
	& = & \frac{4\gamma}{\pi \tOM} \int_0^{\infty} d\omega \int_0^t ds e^{-\gamma(t-s)} \sin \tOM (t - s) \w \cos\w(t-s)\label{eq:otwosource}
\eea
Defining $\w\cos\w(t-s) = - \p_s \sin\w(t-s)$ and integrating by parts, (\ref{eq:otwosource}) becomes
\bea
	F(t)  & = & \frac{4\gamma}{\pi\tOM} \int_0^{\infty}d\w \int_0^t ds e^{-\gamma(t - s)} \sin\w(t-s) \nonumber\\
	& & \qquad \qquad \times \left(\gamma \sin\tOM(t-s) - \tOM \cos\tOM(t-s)\right)\nonumber\\
	& = & \frac{4\gamma}{\pi\tOM} \int_0^{\infty} \frac{d\omega}{\left(\gamma^2 + (\tOM - \omega)^2\right)\left(\gamma^2 + (\tOM + \omega)^2\right)}\nonumber\\
	& & \qquad \qquad \times \left\{ \tOM\omega(\gamma^2 + \tOM^2 - \omega^2) \right.\nonumber\\
	& & \qquad \qquad +  e^{-\gamma t} \cos(\tOM t) \tOM \omega\left((\omega^2 - \gamma^2 - \tOM^2)\cos\omega t + 2\gamma\omega\sin\omega t\right) \nonumber\\
	& & \qquad \qquad  - e^{-\gamma t} \sin (\tOM t) \left(\gamma\omega(\gamma^2 + \tOM^2 + \omega^2)\cos\omega t \right. \nonumber\\
	& & \qquad \qquad\qquad \qquad \left. \left. + [(\gamma^2 + \tOM^2)^2 + (\gamma^2 - \tOM^2)\omega^2]\sin\omega t\right)\right\}\label{eq:factforce}
\eea
Focusing on the large-$\omega$ part of the integral (thus, for example, taking the denominator to scale as $\omega^4$), we find:
\be
	F_{div} = \frac{4\gamma}{\pi \tOM} \int^{\infty} \frac{d\omega}{\omega} \left[ - \tOM + e^{-\gamma t} \left(\tOM \cos\tOM t - \gamma \sin \tOM t \right)\cos\omega t\right]\label{eq:divforce}
\ee

The constant logarithmic divergence in (\ref{eq:divforce})\ is equal and opposite to the logarithmic divergence in $2\vev{O_3}/M$, both in the ground state and in this factorized initial state, as we can see from the results in Appendix A.  If we write 
\be
	F_{log} = - \frac{4\gamma}{\pi} \int^{\Lambda}_{\eps} \frac{d\omega}{\omega} = - \frac{4\gamma}{\pi}\ln\Lambda/\eps
\ee
and define
\be
	O_{3,r} = O_3 - F_{log} 
\ee
as the renormalized operator $P^2_r$ with finite matrix elements, we find:
\bea
	\frac{d}{dt} \vev{O_2} & = & \frac{2}{M} \vev{O_{3,r}} - 2 \left(M \Omega_r^2 + 8\gamma M\delta(t)\right) \left(\vev{O_1} - X_0^2\right) \nonumber\\
	& & \qquad \qquad - 2\gamma \vev{O_2}+ F(t) - F_{log}
\eea
Now $F_{div} - F_{log}$ is logarithmically divergence in the limit $t \to 0$.  In fact, the second term in (\ref{eq:divforce}) cancels the constant divergent term at $t = 0$: however, the cancellation fails for $t > 0$, and $F_{div} - F_{log}$ dies off exponentially in $t$.  Since $F_{div} - F_{log}$ has the time dependence characteristic of the UV oscillators, we believe that this term is the result of the initial state containing high-energy modes. 

In \cite{Agon:2014uxa}, we emphasized the importance of understanding ones resolution in time; a finite time resolution means that one should average observed quantities over the scale of one's resolution.  In the present case, the extra time-dependent parts in (\ref{eq:factforce}) all oscillate with the frequency of the environment oscillators.  We have allowed those frequencies to extend to zero for ease of calculation.  However, if we impose a lower cutoff $\omega_{min}$ to the environment oscillators, the divergence structure of our results will not change. Meanwhile, if we average $F(t)$ over a time scale $\tau_{res} \gg \omega_{min}^{-1}$, the time-dependent parts of $F_{div}$ will vanish exponentially in $\tau_{div}\omega_{min}$. To good approximation, then, we can set the time average ${\overline F}_{div} = F_{log}$.

\subsection{Born-Oppenheimer states}

States such as (\ref{eq:factinitBO})\ remove the initial jolt in one-point functions, as described in \cite{RosenaudaCosta:1999uj}.  Because the full system is quadatic, the evolution of the one-point functions depends only on their initial values and not on any further structure of the wavefunction such as the width.  In this sense the authors of \cite{RosenaudaCosta:1999uj} are correct that entanglement between system and environment is not required to remove the initial jolt from the equation for $\vev{X}$, $\vev{P}$.  

The quadratic operators $X^2$, $P^2$, and $\{X,P\}$, however, certainly do depend on this structure, and as we have seen the condition laid down in \cite{RosenaudaCosta:1999uj}\ is not sufficient to remove the intial jolt in the expectation values of these operators.  

To find states without such a jolt, we take a preliminary step at projecting out high-energy states. Inspired by the wavefunction (\ref{eq:factinitBO}) in which the environment oscillators are in an instantaneous ground state centered on $\vev{X}$, we choose an initial quantum state following the lowest order Born-Oppenheimer approximation.  Under the assumption that the UV oscillators $x_i$ have frequencies $\omega_i \gg {\tilde \Omega}$, we consider the state
\be
	\ket{\Psi} = \int dX \Psi(X) \ket{X}_{IR} \ket{0;X}_{UV}\label{eq:zobo}
\ee
Here $\ket{0;X}$ denotes the state of the oscillators $x_i$, if they were described by simple harmonic oscillators with frequency $\omega_i$ and mass $m_i$, centered at the position $\frac{C_i}{m_i\omega_i^2} X$.  Note that these states {\it are}\ entangled between the UV and IR \cite{Grover:2013tia,Agon:2014uxa}.\footnote{Ref. \cite{Grover:2013tia}\ uses the entanglement properties of Born-Oppenheimer-like wavefunctions to find interesting entanglement properties of two-component fluids: the goals and results are distinct from the discussion here.}

We will be a bit lazy on two points. First, by taking (\ref{eq:zobo}) as the initial state to use in solving the equations in \S2.3, we will remove the delta-function initial jolts seen there, but we will be left with additional high-frequency transients just as we did for the factorized initial states.  This is presumably because the equations have ${\cal{O}}(\gamma)$ terms: to project out high-frequency states to ${\cal{O}}(\gamma)$ in (\ref{eq:zobo}) will require working to the subleading order in the Born-Oppenheimer approximation. We feel a systematic treatment of the Caldeira-Leggett model in this approximation, and more generally a computation of the master equation for states constructed via the Born-Oppenheimer approximation, are interesting projects in and of themselves, and we present this section in part as motivation for future work. 

Secondly, we will use the state (\ref{eq:zobo})\ for all of the environmental oscillators. The Born-Oppenheimer approximation works for $\omega_i \gg \Omega$, but as we are for convenience extending the envornment down to $\omega_i \to 0$, we will keep the form (\ref{eq:zobo})\ for these as well.  In practice we expect these low-frequency oscillators to get excited above their instantaneous ground state, via the perturbative coupling, at a rate higher than the high-frequency oscillators.  However, as we are focused on the divergence structure due to high-frequency modes, we will not worry about this. At any rate, (\ref{eq:zobo})\ is a perfectly legitimate initial state and our calculations are valid at every step.

Let us again focus on the equation (\ref{eq:dtpp}).  As before, we use (\ref{eq:fullheisx}) to write the final term entirely in terms of the environmental oscillators. The full computation of the resulting equation is a bit long, so we leave the details for Appendix \ref{B}.
%
%
%
The result is the equation
\begin{eqnarray}
	\frac{d}{dt} \vev{O_2} & = & \frac{2}{M} \vev{O_3} - 2 M \Omega_r^2 \vev{O_1} - 2\gamma \vev{O_2} 
		- 4\gamma G(t) e^{-\gamma t} \frac{\sin{\tilde \Omega} t}{{\tilde \Omega}}\nonumber\\
	& & + \sum_{j,k} C_j C_k \int ds e^{-\gamma(t-s)} \frac{\sin {\tilde \Omega}(t-s)}{M{\tilde \Omega}} \vev{\{x^j_0(s),x^k_0(s)\}}\label{eq:redOtwo}
\end{eqnarray}
where 
\be
	G(t) = \int_0^{\infty} \frac{d\omega}{\pi} \sin \omega t
\ee
must be defined by cutting off the frequency integral.

Evaluating the last line in the Born-Oppenheimer state, as detailed in Appendix B, yields a term cancelling  the fourth term on the RHS of (\ref{eq:redOtwo}), leading to:
\begin{eqnarray}
	\frac{d}{dt} \vev{O_2} & = & \frac{2}{M} \vev{O_3} - 2 M \Omega_r^2 \vev{O_1} - 2\gamma \vev{O_2} 
		\nonumber\\
	& & - \frac{4\gamma}{{\tilde \Omega}} \int_0^{\infty} \frac{d\omega}{\pi} \int_0^t ds \sin\omega(t-s) e^{-\gamma(t-s)}\left(\gamma \sin{\tilde \Omega}(t-s) - \right.\nonumber\\
	& & \qquad \qquad \qquad \qquad 
	\left. {\tilde \Omega}\cos{\tilde \Omega}(t-s)\right)\label{eq:boxp}
\end{eqnarray}
This equation is essentially identical to (\ref{eq:dtxpfact}), only with the delta function jolt term completely cancelled.  As in that case, there is a constant divergent part in the last term of (\ref{eq:boxp}) which cancels the constant divergent part of the unrenormalized operator $\vev{O_3}$.  
The time-dependent of $F_{div}$ in (\ref{eq:divforce}), which diverge as $t \to 0$, remain as before.
If we assume that the frequency spectrum for $x_i$ starts at high energies, the second term notably contains the $\cos\omega t$ dependence indicating that the lowest-order Born-Oppenheimer state contains admixtures of high-energy excitations at order ${\cal O}(\gamma)$. We expect that we can construct a corrected state of the form
\be
	\ket{\Psi} = \int dX  \left[ \Psi(X) \ket{X}_{IR} \ket{0;X}_{UV} + \sum_n \delta \Psi_n(X) \ket{X}_{IR} 
		\ket{n; X}_{UV} \right]
\ee
in which the system is projected onto the subspace consisting of energies lower than those of the UV oscillators, so that the rapidly oscillating terms in (\ref{eq:divforce}) would be cancelled.  The further study of the Born-Oppenheimer approximation in this and related models is clearly an important topic for future work.


\subsection{Master equation for factorized initial states}

\subsubsection{Setup}

Finally, we turn to the main point of this section, the description of the oscillator $X,P$ as an open quantum system, and the computation of the associated master equation 
\be
	i \p_t \rho(t) = {\cal L}[\rho]
\ee
for the reduced density matrix 
\be
	\rho = {\rm tr}_{x_i} \ket{\Psi(t)}\bra{\Psi(t)}
\ee

There is at present no complete general theory for the structure of ${\cal L}$.  The best-known class of master equations \cite{Kossakowski1972247,Lindblad:1975ef}, describes quantum Markov processes:
\bea
	i \p_t \rho & =  &[H,\rho] + i \gamma[\rho] - i \{A,\rho\} \nonumber\\
	& = & [H_{eff}, \rho] + i \sum_k h_k \left[L_k \rho L_k^{\dagger} - \half \{L_k^{\dagger} L_k, \rho \}\right]
	\label{eq:klmaster}
\eea
Here $h_k > 0$ , and $L_k$ is an arbitrary basis of operators. $H_{eff}$ includes the part of the full Hamiltonian acting solely on the observed subsystem, as well as effects from virtual processes involving the unobserved degrees of freedom which will renormalize the Hamiltonian. Markovian evolution is of course not the most general case. If $\rho$ is the reduced density matrix for a factor ${\cal H}_{obs}$ of the Hilbert space ${\cal H} = {\cal H}_{obs}\otimes{\cal H}_{unobs}$ of a closed quantum system, then the unobserved factor can carry memory.  One must make a specific set of assumptions and approximations for the master equation to be Markovian: we will not do so here.

For states with vanishing entanglement at time $t = 0$ between the observed (system) and unobserved (environment) factors of the Hilbert space, the master equation generically has a nearly identical structure:
\bea
	i \p_t \rho & =  &[H,\rho] + i \gamma[\rho] - i \{A,\rho\} \nonumber\\
	& = & [H_{eff}, \rho] + i \sum_k h_k(t) \left[L_k \rho L_k^{\dagger} - \half \{L_k^{\dagger} L_k, \rho \}\right]
	\label{eq:initfactmaster}
\eea
where, however, $h_k(t)$ can be time-dependent and nonpositive (see \cite{breuer2007theory,BreuerFoundations:2012,2014RPPh...77i4001R}\ for reviews and references.) We will consider this case here and in \S3.

Ref \cite{Agon:2014uxa}\ examined the detailed structure of $H_{eff},\gamma, A$ for the case that ${\cal H} = {\cal H}_{IR} \otimes {\cal H}_{UV}$; the Hamiltonian has the structure
\be
	H = H_{IR}\otimes {\bf 1}_{UV} + {\bf 1}_{IR} \otimes H_{UV} + \lambda V\ ;
\ee
and the spectra of $H_{IR},H_{UV}$ is characterized by energy gaps $E_{IR} \ll E_{UV}$. That work treated the $\lambda V$ term as a perturbation and computed the master equation to second order in $\lambda$, before and after a physical time-averaging procedure was applied.  The dynamics of the master equation for the reduced density matrix $\rho_{IR}$ in ${\cal H}_{IR}$ was taken to be a model for coarse-grained dynamics of quantum field theories.  As we will see in \S3, when we apply the formalism of \cite{Agon:2014uxa}\ to a scalar quantum field theory in general spacetime dimensions, using a Hamiltonian regularization scheme, the terms $\gamma,A$ have divergences in spatial dimensions higher than three. Understanding this is the motivation for this section.

Similarly, we will see that the Caldeira-Leggett model with Ohmic dissipation has lo-garithmic divergences in the non-Hamiltonian part of the master equation for the ``system" (with observables $X$,$P$) that are of the same form as scalar field theory with cubic self-interactions in four spatial dimensions.  Now that we have explored the structure of this model, we can turn to the direct calculation of these divergences and interpret them.

%
 

\subsubsection{Perturbative computation}

We begin with the full system in the state 
\be
	\sigma(0)=|\bar{0}\rangle \langle \bar{0}|\otimes \rho(0)\ ,\label{eq:factdens}
\ee
with $\rho(0)$ the initial density matrix for the IR degrees of freedom, and $|{\bar 0}\rangle$ the ground state of the environment variables when $C_j = 0$.  It would be interesting to work out the master equation for the ``Born-Oppenheimer" states we studied in \S2.5, but we leave this for future work: as far as we know, a general form for the master equation for such states has not been worked out.\footnote{However, the dissipative classical dynamics of ``heavy" particles coupled to an environment of light quantum particles {\it have}\ been calculated in the Born-Oppenheimer approximation in \cite{2014AnPhy.345..141D,2014arXiv1405.2077D}.}   We have not chosen the ``shifted" factorized states in which the initial jolt is subtracted from the Heisenberg equation for the operators linear in $X,P$, as we are interested in seeing how this initial jolt appears in the master equation; furthermore, our experience in prior sections convinces us that we can separate out this initial jolt from time-independent divergences that survive projecting out high-energy components of the initial state.


We will consider the master equation to second order in perturbation theory, following  \cite{Agon:2014uxa}.  The full theory is Gaussian, but our goal is to cast light on the perturbative results for self-interacting field theories for which no exact analysis is available. As reviewed in \cite{Agon:2014uxa}, the terms in the master equation have the structure:
\bea
H_{eff} & = & \frac{P^2}{2 M}+ \frac{1}{2} M\W^2 X^2+H^{(2)}(t)\nonumber\\
A(t) & =& A^{(2)}(t) \nonumber\\
\gamma(t)& = & \gamma^{(2)}(t)\,,
\eea
where the superscript $(2)$ denotes the terms that are second order in perturbation theory.  Applying that formalism to   (\ref{sys}), with the term $- \sum_j C_j x_j X$ taken as the interaction term, we find that:
%
\begin{eqnarray}
H^{(2)}(t)&=&-\frac i2\sum_j  \frac{C^2_j}{2m_j \omega_j}  \Big[\int_0^t d\tau e^{-i\w_j\tau}  X \cdot  X(-\tau) -\int_0^t d\tau e^{i\w_j\tau}  X (-\tau)\cdot  X\Big]\,,\nonumber
\\
iA^{(2)}(t)&=&-\frac i2\sum_j  \frac{C^2_j}{2m_j \omega_j}  \Big[\int_0^t d\tau e^{-i\w_j\tau}  X \cdot  X(-\tau) +\int_0^t d\tau e^{i\w_j\tau}  X (-\tau)\cdot  X\Big]\,,\nonumber\\
\end{eqnarray}
and 
\begin{eqnarray}
\gamma^{(2)}(t)&=
& i \sum_j  \frac{C^2_j}{2m_j \omega_j}  \left[\int_0^t d\tau e^{-i\w_j\tau}  X(-\tau) \rho^{(0)}(t) X \right.\nonumber\\
& & \qquad \qquad \left. +\int_0^t d\tau e^{i\w_j\tau}  X \rho^{(0)}(t) X (-\tau) \right]\,. \nonumber\\ 
\end{eqnarray}
where $X$ without an argument is shorthand for the Schr\"odinger picture operator, and $X(-\tau)$ denotes the associated interaction picture operator.

Choosing the Ohmic spectrum of oscillators, and defining 
\be
	C(\tau)\equiv \int_0^\infty \frac{d\w}{\pi }\, \w e^{-i\w \tau}\ ,
\ee
we find:
\begin{eqnarray} \label{FXX}
H^{(2)}(t)&=&- i\gamma M \int_0^t d\tau\Big[C(\tau)  X \cdot  X(-\tau) -C^{*}(\tau) X (-\tau)\cdot  X\Big]\,,\nonumber
\\
iA^{(2)}(t)&=&- i\gamma M \int_0^t d\tau  \Big[ C(\tau)  X \cdot  X(-\tau) +C^{*}(\tau) X (-\tau)\cdot  X\Big]\,,\nonumber\\
\gamma^{(2)}(t)&=
&+2i \gamma M \int_0^t d\tau \Big[ C(\tau)   X(-\tau) \rho^{(0)}(t) X+C^{*}(\tau)  X \rho^{(0)}(t) X (-\tau) \Big]\,.\nonumber\\ 
\label{eq:meqops}
\end{eqnarray}

Next, we write
\be
C(\tau)=\int_0^\infty \frac{d\w}{\pi} \w e^{-i\w \tau}=i\delta'(\tau)+G'(\tau) \label{eq:meqcoeff}
\ee
where 
\be
G'(\tau)=\int_0^\infty \frac{d\w}{\pi} \w \cos\w \tau\,.
\ee
Thus the $\tau$ integrals in (\ref{eq:meqops})\ that involve the first term on the right hand side of (\ref{eq:meqcoeff})\ can be done exactly, and all derive from:  
\bea
\label{Gintegral}
\int_0^{t} d\tau C(\tau)X(-\tau)&=&i \int_0^{t} d\tau \delta'(\tau)X(-\tau)+\int_0^{t} d\tau G'(\tau)X(-\tau)\nonumber \\
&=&i\left(\delta(t)-\frac{\Lambda}{\pi}\right)X_S+\frac{i}{2M}P_S+\int_0^{t} d\tau G'(\tau)X(-\tau)\nonumber \\
\label{eq:meqctwo}
\eea
where the first two terms on the last line arise from an integration by parts of the first term on the right hand side of the previous line; and we use a regularized expression for $\delta(0)=\Lambda/\pi$. 

We can evaluate the last term in (\ref{eq:meqctwo}) exactly, using the hard cutoff (\ref{eq:hardcut}), in terms of sine and cosine integrals. When $\Lambda/\Omega \gg 1$, $\Lambda t \gg 1$, we find:
\bea
	\int_0^t d\tau G'(\tau) X(-\tau) 
	& = & \frac{\Omega}{2} X_S + \frac{1}{\pi M } \left( \ln\frac{\Lambda}{\Omega}\right) P_S + h(t) X_S + k(t) P_S
	\label{eq:meqint}
\eea
Here $h,k$ fall off as powers of $\Lambda t$. 
\footnote{From our experience working with the exact Heisenberg equations in \S2.2,\S2.3, such as in (\ref{eq:factforce}), we might expect the denominators in (\ref{eq:meqint}) to be shifted by a factor of $\gamma$, and for there to be an overall damping factor. The shift and the damping rate in (\ref{eq:factforce}), however, scale as $\gamma$, so we would need to go to higher orders in perturbation theory to see them.}

%

The coefficient of $X$ in (\ref{eq:meqint}), as well as the time-dependent pieces proportional to $h(t)$, $k(t)$, get their main contribution from $\omega \ll t^{-1}$. If we take $t$ large compared to the temporal resolution of our experiment, they will survive the time averaging. That said, $h,k$ vanish as powers of $t$ so we henceforth ignore the explicitly time-dependent parts of (\ref{eq:meqint}), taking $t$ sufficiently large that these terms are small.

Keeping the leading order terms in $\Lambda$, we have:
\bea
\label{Ftt'2}
\int_0^{\infty} d\tau C(\tau)X(-\tau)
&=& \left[\frac{ \W }{2}-\frac{i}{2}\(\frac \Lambda \pi -\delta(t)\)\right] X_S -\left[\frac{\ln(\W/\Lambda)}{\pi M }-\frac{ i}{2M}\right]P_S\nonumber \\
\eea
Plugging this back into (\ref{eq:meqops}), we find:
\begin{eqnarray}
H^{(2)}(t)&=&-2\gamma M \left(\frac{\Lambda}{\pi}-\delta(t) \right) X^2 -\frac{\gamma}{\pi}\ln\left(\frac \W\Lambda \right)+\frac \gamma 2 (XP+PX)\,,\nonumber
\\
iA^{(2)}(t)&=&-i\gamma M\W X^2   +i\frac{\gamma}{\pi}\ln\left(\frac \W\Lambda \right) (XP+PX) + \frac{i\gamma}2 \,,\nonumber\\
\gamma^{(2)}(t)&=
&+2i \gamma M \W X\rho^{(0)}(t) X-2 i\frac{\gamma}{\pi}\ln\left(\frac \W\Lambda \right) (P\rho^{(0)}(t)X+X\rho^{(0)}(t)P) \nonumber\\ 
&&-\gamma(P\rho^{(0)}(t)X-X\rho^{(0)}(t)P)\,, \label{eq:tavmast}
\end{eqnarray}
The linear divergence in $H^{(2)}$ renormalizes the frequency: to this order in perturbation theory, $\Omega_r^2 = \Omega^2 - \frac{4\gamma M \Lambda}{\pi}$, which matches the exact renormalization above.  In addition, there is
a delta function ``jolt" in the Hamiltonian, appearing as a time-dependent part of the frequency of the renormalized system oscillator.  As we will see this reproduces the equations of motion for expectation values of the system operators.  Finally, aside from an unobservable constant in the Hamiltonian, there are logarithmic divergences in the non-Hamiltonian part of the master equation.  These lead to the logarithmic divergences we found in the Heisenberg equations of motion, when the master equation is used to compute the equation of motion for the quadratic operators $\vev{O_k}$.

We can write the master equation using the operators (\ref{eq:tavmast})\ in the form: 
\bea
\label{master}
\frac{d}{dt}\rho(t)&=&\frac{1}{i}\left[H_{eff},\rho(t) \right]-\gamma M\W [X,[X,\rho(t)]] \nonumber \\
&& +\frac{\gamma}{i} ([X,\rho(t)P]-[P,\rho(t)X])\nonumber\\
& &+\frac{\gamma}{\pi}\ln\left(\frac{\W}{\Lambda}\right)([X,[P,\rho(t)]]+[P,[X,\rho(t)]])\,. 
\eea
This differs from the finite $T$ expression obtained in \cite{Caldeira:1982iu}: our expression contains an extra term with a large logarithmic coefficient $\sim \ln(\W/\Lambda)$, and the coefficient of the term $[X,[X,\rho(t)]]$ present in both cases differs in the two expressions.  This is a result of the particular limit used in \cite{Caldeira:1982iu}, for which $T \gg \Lambda$, as is discussed carefully in \cite{breuer2007theory}. In Appendix \ref{LargeT}, will discuss the comparison of these models in some detail.

\subsection{Time evolution of IR observables}

The logarithmic divergence found in (\ref{eq:tavmast}), and its analog for quantum field theories described in \S3, were the impetus of this work.  However, in computing the exact Heisenberg equations for $X$, $P$, and the quadratic operators $O_k$, we also found explicit divergences which could be interpreted as renormalizing the Hamiltonian as well as the composite operator $O_3 = P^2$.  To tie these discussions together, we now show that the divergences in (\ref{eq:tavmast}) lead precisely (at the order of perturbation theory in which we work) to the divergences we found earlier in the Heisenberg equations of motion.
%

Given the master equation  (\ref{master}), we can compute the equation for the time evolution of any operator ${\cal O}$ acting on the ``system" Hilbert space:
\bea
\frac{d}{dt}\langle {\cal{O}}\rangle&=&\frac{1}{i}\langle \left[{\cal{O}},H_{ren}\right]\rangle+\frac{\gamma}{2i}\langle [{\cal{O}},\{X,P\}]\rangle-\gamma M\W \langle [X,[X,{\cal{O}}]]\rangle \nonumber \\
&& +\frac{\gamma}{i} \left(\langle X[P,{\cal{O}}]\rangle -\langle P[X,{\cal{O}}]\rangle \right)\nonumber\\
& & +\frac{\gamma}{\pi}\ln\left(\frac{\W}{\Lambda}\right)(\langle [P,[X,{\cal{O}}]]\rangle +\langle [X,[P,{\cal{O}}]]\rangle)\,, \nonumber \\
\eea
where $H_{ren}$ is equal to the original IR Hamiltonian but with $\W^2 \to {\W}_r^2+4\gamma \delta(t) $, where ${\W}^2_r=\W^2-4\gamma \Lambda/\pi$ is the renormalized frequency.

The resulting equations for linear and quadratic operators are:
\bea
\label{evo}
\frac{d}{dt}\langle X \rangle &=&\langle P \rangle/M \nonumber \\
\frac{d}{dt}\langle P \rangle &=&-M({\W}^2_r+4\gamma \delta(t)) \langle X \rangle -2\gamma \langle P\rangle \nonumber \\
\frac{d}{dt}\langle O_1 \rangle &=&\frac{\langle O_2\rangle}{M}\nonumber \\
\frac{d}{dt}\langle O_2 \rangle &=&\frac2M\langle O_3 \rangle-2M({\W}_r^2+4\gamma \delta(t)) \langle O_1\rangle -2\gamma\langle O_2\rangle + \frac{4 \gamma}{\pi} \ln\left( \W / \Lambda \right) \nonumber \\
\frac{d}{dt}\langle O_3 \rangle &=&-M({\W}_r^2+4\gamma \delta(t)) \langle O_2\rangle -4\gamma \langle O_3\rangle +2\gamma M \W \nonumber \\
\eea
This matches the divergence structure of the equations in \S2.3\ to second order in $C_i$; when we perform the time averaging of those equations, the two derivations give exactly the same expressions to this order. The logarithmic divergence in the equation for $O_2$, which combines with $P^2$ to form a renormalized operator, follows precisely from the logarithmic divergences in the non-Hamiltonian part of the master equation. The divergence in the master equation 
is thus required for the Heisenberg equation of motion to close on renormalized operators.

%
%
%

\section{Scalar quantum field theory}

We now proceed to our original goal of understanding the divergence structure of the master equation for coarse-grained quantum field theories. Following the discussion in \cite{Agon:2014uxa}, we consider a situation in which our experimental apparatus can only read or manipulate degrees of freedom with spatial momenta larger than some scale $\Lambda \ll \Lambda_0$, where $\Lambda_0$ is the fundamental cutoff on spatial momenta in the theory.

As we will see, the non-Hamiltonian parts of the master equation for factorized initial states have a similar divergence structure in spatial dimensions higher than three. As in the Caldeira-Leggett model, these divergences ensure that the equations of motion for composite operators close on renormalized operators with finite matrix elements. To motivate this, we compute the {\it vacuum}\ expectation values of the composite operators which appear in these equations. This has a divergence which precisely cancels the explicit divergence coming from the divergent term in the master equation, as does the expectation value of the same combination in the initially factorized state actually used to construct the reduced density matrix.  

Interestingly, the individual terms in this combination have different divergence structures in the two different states, reflecting the fact that the factorized state has high-energy short-wavelength excitations with respect to the actual vacuum state.  This is reminiscent of the change in the divergence structure for composite operators for $\alpha$-vacua in de Sitter space: see \cite{Kaloper:2002cs}\ and references therein. 

\subsection{Background and notation}

We consider $\phi^3$ theory in $d$ spatial dimensions, with Lagrangian density
\bea
\mathcal{L}=\frac 12 \p^\mu \phi \p_\mu\phi -\frac 12 m^2 \phi^2-\frac \lambda {3!} \phi^3\,.
\eea
The Hamiltonian is divided in free and interactive parts $H=H_0+V$ 
\bea
H_0=\int d^dx (\frac 12 \pi^2+\frac{1}{2}(\nabla \phi)^2+ \frac 12 m^2 \phi^2) \quad {\textrm{and }} \quad
V=\frac \lambda {3!} \int d^dx \phi^3\,,
\eea
where $\pi=\partial_t \phi$ is the canonical conjugate momenta to the field $\phi$, and $d$ is the number of spatial dimensions. This theory is, of course, nonperturbatively unstable, but we will focus on perturbation theory and ignore that issue here.

We adopt the following conventions:
\begin{itemize}
\item The field expansion in the interaction picture is 
\bea
\phi(x,t)=\int \frac{d^dp}{\sqrt{(2\pi)^d 2\w_p}}(e^{-i\w_p t}a_p+e^{i\w_p t}a^{\dagger}_{-p})e^{ip\cdot x}
\eea
where $\omega_k = \sqrt{{\vec k^2} + m^2}$

\item Momentum eigenstates are $|k\rangle=\sqrt{2 \w_k}a^{\dagger}_k|0\rangle$ where the $a_k's$ satisfy the usual commutation relations $\left[a_k,a^\dagger_{k'}\right]=\delta^{d}(k-k')$.

\item  We take infrared (IR) modes with spatial momenta $0< |{\vec k}| <\Lambda$ to be observable, the analog of the ``system" coordinates in the Caldeira-Leggett model. Ultraviolet (UV) modes with spatial momenta $\Lambda< |{\vec k}| <\Lambda_0$ are taken to be ``environment" modes. We can then write $\phi=\phi_{ir}+\phi_{uv}$ in an obvious fashion.

\item With the normalization given above, the identity operator on the space of UV modes can be represented as:
\bea
\mathbf{1}=|0\rangle \langle 0|+\int_{uv}\frac{d^dk}{2\w_k}|k\rangle\langle k|+\frac1{2!}\int_{uv} \frac{d^dk}{2\w_k}\frac{d^dk'}{2\w_{k'}}|k k'\rangle\langle k k'|+\cdots\ ,
\eea
an expression we will have cause to use below.

\end{itemize}


Note that we will coarse-grain {\it spatial}\ momenta below the scale $\Lambda$, and define the theory with a fundamental {\it spatial}\ cutoff $\Lambda_0$, as is natural for the Hamiltonian treatment we adopt here -- {\it cf}\ \cite{Kogut:1974ag}. 
Temporal coarse-graining at the scale $\Lambda/c$ is implemented via a physical time-averaging procedure, following the discussion in \cite{Agon:2014uxa}.  

\subsection{Computing the master equation to order ${\cal O}(\lambda^2)$}

We start by computing the master equation for factorized initial states $\ket{\psi}_{ir} \ket{0}_{uv}$, where $\ket{\psi}$ is 
an arbitrary state constructed from oscillators with $|{\vec k}| < \Lambda$ and $\ket{0}_{uv}$ is the noninteracting vacuum for the UV oscillators.

Written in terms of $\phi_{ir,uv}$, the free and interacting terms in the Hamiltonian are:
\bea
H_0&=&\int d^dx (\frac 12 \pi_{ir}^2+\frac{1}{2}(\nabla \phi_{ir})^2+ \frac 12 m^2 \phi_{ir}^2) 
+  ({ ir}\to {uv}) \ , \\
\quad
V&=&\frac \lambda {3!} \int d^dx (\phi_{ir}^3+3\,\phi^2_{ir}\phi_{uv}+3\,\phi_{ir}\phi^2_{uv}+\phi_{uv}^3)\,,
\eea
Both $H_0$ and $V$ should be thought in as normal ordered operators: for simplicity of notation we avoid using the standard $:{\cal O}:$ notation. Note a slight difference from \cite{Agon:2014uxa}, in that we include self-interactions of the UV and IR modes in the perturbation $V$.  We can nonetheless apply the formulae given in that work to construct the master equation to ${\cal O}(\lambda^2)$.

On general grounds the master equation takes the form (\ref{eq:initfactmaster}), with order ${\cal O}(\lambda)$ contribution $H^{(1)}$ to the Hamiltonian and ${\cal O}(\lambda^2)$ contributions $H^{(2)}$, $A^{(2)}$, $\gamma^{(2)}$ to $H$, $A$, and $\gamma$:
\bea
H^{(1)}(t)&=&\frac \lambda {3!} \int d^dx \, \phi_{ir}^3\,, \\
\label{h2a}
H^{(2)}(t)&=&-\frac i2 \int_0^t d\tau \int_{uv} \frac{d^dk}{2\w_k}\left[\langle 0|V|k\rangle\langle k|V(-\tau)|0\rangle -\langle 0|V(-\tau)|k\rangle\langle k|V|0\rangle \right]\nonumber \\
&&-\frac i4 \int_0^t d\tau \int_{uv} \frac{d^dk}{2\w_k}\frac{d^dk'}{2\w_{k'}}\Big[\langle 0|V|k k'\rangle\langle k k'|V(-\tau)|0\rangle\nonumber \\
&& \qquad \qquad \qquad \qquad \qquad \qquad -\langle 0|V(-\tau)|k k'\rangle\langle k k'|V|0\rangle \Big]\nonumber\\
A^{(2)} & = & - \half \int_0^t d\tau \int_{uv} \frac{d^dk}{2\w_k}\left[\langle 0|V|k\rangle\langle k|V(-\tau)|0\rangle + \langle 0|V(-\tau)|k\rangle\langle k|V|0\rangle \right]\nonumber \\
&&-\frac{1}{4} \int_0^t d\tau \int_{uv} \frac{d^dk}{2\w_k}\frac{d^dk'}{2\w_{k'}}\Big[\langle 0|V|k k'\rangle\langle k k'|V(-\tau)|0\rangle\nonumber \\
&& \qquad \qquad \qquad \qquad \qquad \qquad + \langle 0|V(-\tau)|k k'\rangle\langle k k'|V|0\rangle \Big]\nonumber\\
\gamma^{(2)} & = & i \int_0^t d\tau \int_{uv} \frac{d^dk}{2\w_k}\left[ \langle k|V(-\tau)|0\rangle \rho^{(0)}\langle 0|V|k\rangle
+ \langle k|V|0\rangle \rho^{(0)} \langle 0|V(-\tau)|k\rangle \right]\nonumber \\
& & + \frac{i}{2} \int_0^t d\tau \int_{uv} \frac{d^dk}{2\w_k}\frac{d^dk'}{2\w_{k'}}
\Big[\langle k k'|V(-\tau)|0\rangle \rho^{(0)}\langle 0|V|k k'\rangle\nonumber \\
&& \qquad \qquad \qquad \qquad \qquad \qquad + \langle k k'|V|0\rangle \rho^{(0)} \langle 0|V(-\tau)|k k'\rangle \Big]
\label{eq:qftmaster}
\eea 
where $\rho^{(0)}$ is the density matrix evolved by $H^{(0)}$. We have dropped terms with three intermediate UV modes. At this order, they will lead to $c$-numbers which do not contribute to the master equation.

We will focus in this work on divergent terms in (\ref{eq:qftmaster}).  
The terms in (\ref{h2a}) involving a single intermediate UV particle are not divergent: kinematically, a single UV mode must combine with 2 IR modes, so the only momenta contributing will be close to the resolution $\Lambda \ll \Lambda_0$ of the detector. The terms with two UV momenta are a different story. So long as they are nearly back-to-back, so that $|{\vec k} + {\vec k}'| < \Lambda$, they will contribute.  There is a large phase space of such momenta, and they will contribute divergences to the master equation in sufficiently high dimension. We will therefore focus on these.

 
Before any time averaging, (\ref{h2a}) contains terms of the form: 
\be
	\int_0^t d\tau e^{-i(\w_k+\w_{k'} \pm \w_p)\tau} =  -i\frac{1-e^{-i(\w_k+\w_{k'} \pm \w_p)t}}{\w_k+\w_{k'}\pm \w_p} 
	\label{eq:timeinteg}
\ee
We will follow \cite{Agon:2014uxa}\ and time-average the master equation over a scale $\tau\gg \w_{k,k'}^{-1}$, so that we can drop the second time-dependent part of the numerator in (\ref{eq:timeinteg}).  We will also study the equations at $t > 0$, so we are blind to any initial ``jolt" that might appear in the master equation.

We find:
\bea
& & \int_0^t d\tau \,_{uv}\langle k k'|V(-\tau)|0\rangle_{uv} \to_{time\ average} \nonumber\\
%
%
& &\qquad \qquad  -i\lambda\int_{ir} \frac{d^dp}{\sqrt{(2\pi)^d 2\w_p}}\delta^d (p-k-k') \left[ (a_p+a^{\dagger}_{-p})\frac{(\w_k+\w_{k'})}{(\w_k+\w_{k'})^2-\w_p^2} \right.\nonumber\\
& & \quad \qquad \qquad \qquad \qquad \left. + (a^{\dagger}_{-p}-a_p)\frac{\w_p}{(\w_k+\w_{k'})^2-\w_p^2}\right]\Theta(\Lambda- |k+k'|)\ .
\eea
In the meantime,
\bea
\,_{uv}\langle 0|V| k k'\rangle_{uv}=\lambda\int \frac{d^dx'}{(2\pi)^d}e^{i(k+k')\cdot x'}\phi_{ir}(x')
\eea

Combining these, the time averaging of the term:
\be
 \Phi(t) = \int_0^t d\tau \int_{uv} \frac{d^dk}{2\w_k}\frac{d^dk'}{2\w_{k'}}\langle 0|V|k k'\rangle\langle k k'|V(-\tau)|0\rangle
 \ee
 is
\bea
{\overline \Phi}(t) &=&-i\lambda^2\int \frac{d^dx}{(2\pi)^d}\phi_{ir}(x)\int_{ir} \frac{d^dp}{\sqrt{(2\pi)^d 2\w_p}}\int_{uv}\frac{d^dk}{4\w_k \w_{p-k}} e^{i p x} \nonumber\\
& & \qquad \qquad \times \frac{\left[(a_p+a^{\dagger}_{-p}) (\w_k + \w_{k'}) + (a^{\dagger}_{-p}-a_p) \w_p \right]}{[(\w_k+\w_{p-k})^2-\w_p^2]} \Theta(\Lambda-|p+k|)\label{eq:tavterm}
%
\eea
In order to identify the leading divergences in the large k regime we study the regime $\w_k^2\gg \w_p^2$ which implies that $\w_{p-k}\approx \w_k$. 
We define:
\bea
\alpha=\frac{1}{8}\int_{uv}\frac{d^dk}{(2\pi)^d}\frac{1}{\w_k^3}, \quad {\rm and }\quad \beta=\frac{1}{16}\int_{uv}\frac{d^dk}{(2\pi)^d}\frac{1}{\w_k^4}
\eea
Note that $\alpha$ is UV finite for $d \leq 2$, has a log divergence for $d = 3$, and a linear divergence for $d = 4$. $\beta$ has a log divergence in $d = 4$ but is UV finite for $d \leq 3$.

To the first two orders in $\Lambda_0$, (\ref{eq:tavterm}) becomes:
\bea
{\overline \Phi} = -i\lambda^2 \alpha \int d^d x \phi_{ir}^2(x)-\lambda^2 \beta \int d^d x \phi_{ir}(x)\pi_{ir}(x)
\eea
which leads to the following terms in the master equation:
\bea
H^{(2)}&\sim &-\frac{\lambda^2\alpha}{2} \int d^d x \phi_{ir}^2(x)-\frac{\lambda^2 \beta}4 \int d^d x \delta^d_{x}(0) + \ldots\nonumber \\
A^{(2)}&\sim&-\frac{\lambda^2 \beta}4\int d^d x (\phi_{ir}(x)\pi_{ir}(x)+\pi_{ir}(x)\phi_{ir}(x)) + \ldots \\
\g^{(2)}&\sim&i\frac{\lambda^2 \beta}2 \int d^d x (\phi_{ir}(x)\rho(t)\pi_{ir}(x)+\pi_{ir}(x)\rho(t)\phi_{ir}(x)) + \ldots
\eea
We will drop the constant term in $H^{(2)}$, which renormalizes the vacuum energy.

The term proportional to $\alpha \,\phi_{ir}^2$ is UV divergent in three and higher spatial dimensions.  This is just the standard mass renormalization, the analog of renormalizing $\Omega$ in \S2. Note that in $d = 6$ we also expect log divergences to renormalize the terms $\pi^2$ and $(\nabla\phi)^2$.  Because we are only cutting off the spatial dimensions, the divergences will occur with different coefficients, leading to a renormalization of the speed of light.\footnote{This is mentioned in a comment in p. 269 of \cite{Symanzik:1976qb}, citing an apparently unpublished paper of Susskind.} We will focus on the leading divergences here, and put this issue aside. 

To understand the divergences in $A^{(2)}$, $\g^{(2)}$, we use the master equation to compute the time evolution of various operators.  As in the case of the Caldeira-Leggett model, the divergent terms in the non-Hamiltonian part of the master equation do not lead to any divergences in the equations of motion for $\vev{\phi}$, $\vev{\pi}$. On the other hand, the time evolution of quadratic operators is sensitive to the divergences in the  master equation.  Following the discussion in \S2, we consider
the operator $\partial_t\langle \phi \pi +\pi\phi\rangle$. Using the master equation, the ${\cal O}(\lambda^2)$ divergences at leading order in $\Lambda_0$ lead to the equation:
\bea
& & \frac{d}{dt}\langle \phi_{ir} \pi_{ir} +\pi_{ir}\phi_{ir}\rangle \nonumber\\
& & \qquad \quad =  2\langle \pi_{ir}^2\rangle-2(m^2-\lambda^2 \alpha)\langle \phi_{ir}^2\rangle+\langle\phi_{ir} \nabla^2 \phi_{ir} +(\nabla^2 \phi_{ir})\phi_{ir}\rangle\nonumber\\
& & \qquad \qquad \qquad -\lambda \langle\phi_{ir}^3\rangle-\lambda^2 \beta\int_{ir}
\frac{d^dp}{(2\pi)^d} + {\rm finite}\nonumber \\ \label{eq:eomfrommaster}
\eea
where the final term arises from the non-Hamiltonian part of the master equation, and the remaining explicit terms arise from the Hamiltonian part. We can see the mass renormalization, $m_r^2 = m^2 - \lambda \alpha + \ldots$ where the dots denote possible subleading divergences.

The terms
\be
	\langle  2 \pi_{ir}^2-2 m_r^2  \phi_{ir}^2+\phi_{ir} \nabla^2 \phi_{ir} +(\nabla^2 \phi_{ir})\phi_{ir}\rangle
\ee
on the right hand side of (\ref{eq:eomfrommaster})\ have UV divergences at order $\lambda^2$.  One of these is proportional to $\lambda^2 \alpha$: following Appendix A we take these to ensure that the wavefunctional of interest depends on the renormalized parameters. The second is proportional to $\lambda^2 \beta$: this is cancelled by the final divergent term in (\ref{eq:eomfrommaster}).  The upshot is that, as with our discussion of the Caldeira-Leggett model, the divergent terms in the master equation ensure (at least to the order we have checked) that the equations of motion for composite operators close on renormalized operators, with coefficients equal to the renormalized couplings.

Before continuing, it is worth noting the difference between this work and \cite{Lombardo:1995fg}. which studies the Feynman-Vernon influence functional for a scalar field theory in three spatial dimensions with quartic self-interactions. First, we are working in arbitrary dimension -- the divergences here only appear in four and five spatial dimensions.  Secondly, they cut off the UV part of the path integral via dimensional regularization, whereas we impose a spatial cutoff after \cite{Kogut:1974ag}\ followed by a physical time-averaging procedure.


\subsection{Composite operators in the vacuum state}

To orient ourselves we consider the composite operator equations of motion for the full theory in the vacuum state. This will lend some intuition for the above divergence by starting with a familiar and well-understood calculation.

It is an easy matter to use the phase space path integral to show that
\bea
	\langle T \left[ {\cal O} {\dot \phi}(x,t) \right] \rangle & = & \langle T\left[ {\cal O} \pi(x,t) + i \frac{\delta O}{\delta\pi(x,t)}
		\right] \rangle\nonumber\\
	\langle T \left[ {\cal O} {\dot \pi}(x,t) \right]\rangle & = & \langle T\left[ {\cal O} \(-\nabla^{2}\phi -m^{2} \phi-\frac{\lambda}{2} \phi^2 \)
		- i  \frac{\delta O}{\delta\pi(x,t)}\right] \rangle\label{eq:pssd}
\eea
where ${\cal O}$ could be some string of composite operators at different spacetime points. Here the operators are understood to be Heisenberg picture operators. The final terms on the right hand side of each line are the usual contact terms.

Now, if we define $\phi \pi + \pi \phi$ by point-splitting, we find that:
\bea
	& & \left(\frac{d}{dt_x} + \frac{d}{dt_y}\right) \langle \pi(x) \phi(y) + \phi(x) \pi(y)\rangle_{y\to x} \nonumber\\
	& & \qquad \qquad = \langle 2\pi(x)\pi(y) - m^2 \phi(x)\phi(y) \nonumber\\
	& & \qquad \qquad \qquad + \phi(x) \nabla^2\phi(y)
		+ (\nabla^2\phi(x))\phi(y)  \nonumber\\
	& & \qquad \qquad \qquad - \half \lambda\phi^2(x)\phi(y) - \half \lambda\phi(x)\phi^2(y)\rangle_{y\to x}\nonumber\\
	\label{eq:coeom}
\eea
Note that the contact terms in (\ref{eq:pssd})\ cancel. As we take $t_x \to t_y$, we can think of the LHS as the time derivative of the vev of the composite operator $\pi\phi + \phi\pi$.

Since we are working in the vacuum state, which is time-translation invariant, the left hand side of (\ref{eq:coeom}) will vanish.  Nonetheless, it will be illuminating to study the divergent contributions to each term on the right hand side, and see how they vanish.

In this work we are interested in observables, including composite operators, built from IR degrees of freedom.  With the above form, we can define $\pi_{ir}\phi_{ir} + \phi_{ir}\pi_{ir}$ by starting with the point-split version and projecting $\phi,\pi$ onto spatial momenta $|{\vec k}| < \Lambda$. In the final two terms on the right hand side of (\ref{eq:coeom}), the projection of $\phi^2$ onto IR momenta will include terms of the form $(\phi_{uv}^2)_{ir}$, that is, terms quadratic in UV oscillators, for which the spatial momenta are nearly back-to-back. If we thus split $\phi = \phi_{ir} + \phi_{uv}$, $\pi = \pi_{ir} + \pi_{uv}$, we are left with the equation.
\bea
	& & \frac{d}{dt} \langle \pi_{ir}\phi_{ir}(x) + \phi_{ir}\pi_{ir}(x)\rangle\nonumber\\
	& & \qquad \qquad = \langle 2\pi^2_{ir}(x) - m^2 \phi^2_{ir}(x) \nonumber\\
	& & \qquad \qquad \qquad + \phi_{ir}\nabla^2\phi_{ir}(x)
		+ (\nabla^2\phi_{ir})\phi_{ir}(x)  \nonumber\\
	& & \qquad \qquad \qquad - \half \lambda\phi_{ir}^3(x) - \lambda\phi_{ir}(\phi_{ir}(x)\phi_{uv}(x))\nonumber\\
	& & \qquad \qquad \qquad - \half \lambda\phi_{ir}(\phi_{uv}(x))^2_{ir}\rangle\ .
	\label{eq:ircoeom}
\eea
To match to the previous section, we study the above terms to order $\lambda^2$, in perturbation theory.  If we write the above operators in the interaction picture, we can expand each term in a power series in the interaction Hamiltonian, as always:
\be
	\vev{{\cal O}} = \bra{0} T\left[ {\cal{O}}_I e^{-i \int_C dt' V_I(t')}\right] \ket{0}/\bra{0} T e^{-i \int_C dt' V_I(t')}\ket{0}
\ee
Here $\ket{0}$ is the noninteracting vacuum, and $C$ is a contour asymptoting to $(1 - i \eps)t$ for large $|t|$, in order to project the state onto the true vacuum. We write the interaction term as
\be
	V = \frac{\lambda}{3!} \left( \phi_{ir}^3 + \phi_{uv}^3\right) + \frac{\lambda}{2} \left( \phi_{ir}^2 \phi_{uv} + \phi_{ir}\phi_{uv}^2\right)
\ee

In calculating the above terms, we will need a cutoff of some kind on the UV degrees of freedom.  If we were interested in the above calculation for its own sake, we might adopt a more symmetric cutoff by rotating to Euclidean space and using dimensional regularization or an $SO(d-1)$-invariant Wilsonian cutoff.  To match the Hamiltonian discussion in \S3.2, however we will place the cutoff on spatial momenta: that is, we will expand $\phi_{uv}$ in oscillators with momenta up to $|{\vec p}| = \Lambda_0$. To this end, we write the Feynman propagators as a function of {\it spatial}\ momenta:
\bea
	\vev{\phi_{ir,uv}(x) \phi_{ir,uv}(y)} & = & 
		\int_{ir,uv} \frac{d^d p}{(2\pi)^d}\frac{e^{ip(x-z)}}{2E_p}\left[\Theta(t-z^0)e^{iE_p(t-z^0)}\right.\nonumber\\
		& & \qquad \qquad \left.+\Theta(z^0-t)e^{-iE_p(t-z^0)} \right] \nonumber \\
		\label{eq:osprop}
\eea
For the two-point function of $\phi_{ir}$, the momentum integral is over the range $|{\vec p}| < \Lambda$. For the two-point function of $\phi_{uv}$, the range is $\Lambda \leq |{\vec p}| \leq \Lambda_0$.

To ${\cal{O}}(\lambda^2)$, the divergences on the right hand side of (\ref{eq:ircoeom}) will come from the contributions of the quadratic terms at second order in the interaction Hamiltonian, and contributions from the $\phi_{ir}\phi_{uv}^2$ term, at first order in the interaction Hamiltonian.  In the former case, the divergence terms will come from the $\phi_{ir}\phi_{uv}^2$ part of $V$.
In this latter case, it is again the explicit factor of $\phi_{ir}\phi_{uv}^2$, contracted against the part of the interaction Hamiltonian with the same structure, that will lead to a divergence. This term is most closely related to the divergence in the master equation, and we will see that it has the same divergence structure.  Let us focus on that term first.

To first order in the interaction Hamiltonian,
\bea
& & \langle \W| T\{\phi_{ir}(x)\phi^2_{uv}(y)\} |\W \rangle \nonumber\\
& & \qquad \qquad = -\frac{ i \lambda}{2}\int_{-\infty(1-i\epsilon)}^{\infty(1-i\epsilon)} dz^0 \int d^dz \langle 0| T\{\phi_{ir}(x)\phi^2_{uv}(y) \phi_{ir}(z)\phi^2_{uv}(z) \} |0 \rangle  \nonumber \\\label{eq:secordcub}
\eea
The limits of the integral act to project $\ket{0}$ onto the true ground state.  
%
Three propagators contribute, one $ir$ propagator and two $uv$ operators. There is a symmetry factor of two from the two identical contractions of the $uv$ fields.  The $z^0$ integral over the product of the terms in brackets in (\ref{eq:osprop}) gives us:
\bea
&& e^{i(E_1+E_2+E_3)t} \int_t^{\infty(1-i\epsilon)} dz^0\, e^{-i(E_1+E_2+E_3)z^0 } \nonumber\\
& & \qquad \qquad + e^{-i(E_1+E_2+E_3)t} \int_{-\infty(1-i\epsilon)}^t dz^0 e^{i(E_1+E_2+E_3)z^0 }\nonumber \\
&&\qquad = \frac{2}{i(E_1+E_2+E_3)}= \frac{2E_1}{i[E_1^2-(E_2+E_3)^2]}-\frac{2(E_2+E_3)}{i[E_1^2-(E_2+E_3)^2]}\,.
\nonumber\\
\eea
Inserting this into (\ref{eq:secordcub})\ the resulting expression over $d^dz$ leads to a delta function in momentum space, which forces $p_3=-p_1-p_2$ and leads to 
\bea
& & \langle \W| T\{\phi_{ir}(x)\phi^2_{uv}(y)\} |\W \rangle\nonumber\\
& & \qquad = -\lambda \int_{ir} \frac{d^d p_1}{(2\pi)^d}\frac{e^{ip_1(x-y)}}{2E_1} \int_{uv} \frac{d^d p_2}{(2\pi)^d}\frac{\Theta(\Lambda-|p_1+p_2|)}{2E_2 2E_3}\nonumber\\
& & \qquad \qquad \qquad \times 
\left(\frac{2(E_2+E_3)}{(E_2+E_3)^2-E_1^2}-\frac{2E_1}{(E_2+E_3)^2-E_1^2}\right) \,.\nonumber \\
\eea
Taking the limit $x \to y$, the final term in (\ref{eq:ircoeom}) becomes:
\bea
& & \lambda^2\int_{ir} \frac{d^d p_1}{(2\pi)^d}\frac{1}{2E_1} \int_{uv} \frac{d^d p_2}{(2\pi)^d}\frac{\Theta(\Lambda-|p_1+p_2|)}{2E_2 2E_3}\nonumber\\
& & \qquad \qquad \times \left(\frac{2(E_2+E_3)}{(E_2+E_3)^2-E_1^2}-\frac{2E_1}{(E_2+E_3)^2-E_1^2}\right) \nonumber \\
\eea
%
Now, since  $E_2\approx E_3 \gg E_1$, we can expand the denominators in $E_1^2/((E_2 + E_3)^2$. The UV divergences will come from the lowest order terms in this expansion.  Keeping only the leading and subleading 
terms, we find that the final term in (\ref{eq:ircoeom})\ is:
\bea
\approx 2\lambda^2 \int_{ir}  \frac{d^d p}{(2\pi)^d}\frac{1}{2E_p} \frac{1}{8}\int_{uv}\frac{d^d k}{(2\pi)^d}\frac{1}{E^3_k}-\lambda^2\frac{d^d p}{(2\pi)^d}\frac{1}{16}\int_{uv}\frac{d^d k}{(2\pi)^d}\frac{1}{E^3_k} + \ldots\ .
\eea
If we further note that
\bea\label{phi2}
\langle \phi^2_{ir}\rangle=\int_{ir}  \frac{d^d p}{(2\pi)^d}\frac{1}{2E_p}
\eea
we find that this term is equal to 
\bea
2\lambda^2 \alpha \langle \phi^2_{ir}\rangle -\lambda^2 \beta \int_{ir}  \frac{d^d p}{(2\pi)^d} 
\eea
where $\alpha$, $\beta$ are precisely as defined in the previous section. Thus, we have:
\bea
\frac{d}{dt}\langle \phi_{ir}\pi_{ir}+ \pi_{ir} \phi_{ir} \rangle &=& 2 \langle\pi_{ir}^2\rangle -2(m^2-\lambda^2\alpha )\langle \phi^2_{ir}\rangle \,\nonumber \\
&& \!\!\!\!\!\! \!\!\!\!\!\! \!\!\!\!\!\! \!\!\!\!\!\!  +\langle\phi_{ir} \nabla^2 \phi_{ir}\rangle  +\langle(\nabla^2 \phi_{ir})\phi_{ir}\rangle 
-\lambda \langle \phi^3_{ir}(x)\rangle -\lambda^2 \beta \int_{ir}  \frac{d^d p}{(2\pi)^d}  \nonumber \\
\label{eq:eomdiv}
\eea 
which is precisely the equation (\ref{eq:eomfrommaster}). For $d \leq 4$ these are the divergent terms.  In $d = 5$ there are additional logarithmic divergences we have not discussed -- these include terms renormalizing $\pi^2$ and
$\pi\nabla^2\phi + (\nabla^2\phi) \phi$.

It is worth noting that the divergence proportional to $\beta$ is the result of the specific cutoff we impose here.  Were we to use dimensional regularization or some $SO(d+1)$-invariant Wilsonian cutoff in Euclidean space, it would not appear.  However, our purpose is to examine the explicitly Hamiltonian discussion in \S3.2, and for this we must adopt the same regularization scheme.

As we stated above, each side of (\ref{eq:ircoeom}) must vanish. In the right-hand side, then, the divergence proportional to $\beta$ we have calculated must cancel an ${\cal{O}}(\lambda^2)$ divergence in the expectation value of
%
%
\bea
{\cal{O}}' \equiv 2 \pi_{ir}^2 -2(m^2-\lambda^2\alpha )\phi^2_{ir} +\phi_{ir} \nabla^2 \phi_{ir}  +(\nabla^2 \phi_{ir})\phi_{ir} 
\label{eq:ourcomposite}
\eea
This latter divergence is a sign that the operator requires renormalization. Let us take up each component of this operator in turn.

We start with 
\bea
\langle\phi_{ir}^2\rangle\Big|_{\lambda^2} =-\frac{\lambda^2}{4} \int d^{d+1}y \int d^{d+1}z \langle 0| T\{\phi_{ir}^2(x)\phi_{ir}(y)\phi^2_{uv}(y) \phi_{ir}(z)\phi^2_{uv}(z) \} |0 \rangle  \nonumber \\
\eea
where the integrals in time are over the same rotated contour discussed above.
%
%
Using Wick's theorem, and defining
\bea
g^{\pm}_1(t-t')\equiv \Theta(t-t')e^{iE_1(t-t')}\pm\Theta(z^0-t)e^{-iE_1(t-t')} 
\eea
we find, after integrating over $y$ and $z$ and solving for the resulting delta functions on momenta, that:
\bea
\langle\phi_{ir}^2\rangle\Big|_{\lambda^2} \!\!&=&\!\!-\frac{\lambda^2}{2}
 \!\!\int_{ir} \frac{d^d p_1}{(2\pi)^d} \frac{1}{(2E_1)^2}\int_{uv} \frac{d^d p_3}{(2\pi)^d}\frac{\Theta(\Lambda-|p_3+p_1|)}{2E_3\, 2E_4}\nonumber \\
 &&\int\!\!\int_{-\infty(1-i\epsilon)}^{\infty(1-i\epsilon)}dy^0\,dz^0 g^+_1(t-y^0)g^+_1(t-z^0)g^+_3(y^0-z^0)g^+_4(y^0-z^0)\nonumber \\
\eea
Performing the time integrals, we find:
\bea
\int_{-\infty(1-i\epsilon)}^{\infty(1-i\epsilon)}dy^0\,dz^0 g^+_1(t-y^0)g^+_2(t-z^0)g^+_3(y^0-z^0)g^+_4(y^0-z^0)\nonumber \\
=-\frac{2}{E_1(E_1+E_3+E_4)}-\frac{2}{(E_1+E_3+E_4)^2}\approx -\frac{2}{E_1(E_3+E_4)}\nonumber \\
\eea
 notice the cancellation of the first subleading term in power of the energy ratio $E_1/(E_3+E_4)$.
The leading term in an expansion of $E_{ir}/E_{uv}$ is then:
\bea
\langle\phi_{ir}^2\rangle\Big|_{\lambda^2}\!\!&=&\!\! \lambda^2
 \!\!\int_{ir} \frac{d^d p_1}{(2\pi)^d} \frac{1}{(2E^2_1)(2E_1)}\int_{uv} \frac{d^d p_3}{(2\pi)^d}\frac{\Theta(\Lambda-|p_3+p_1|)}{2E_3\, 2E_4(E_3+E_4)}
\eea
If we consider the combination 
\bea
 -2m^2\langle \phi^2_{ir}\rangle +\langle\phi_{ir} \nabla^2 \phi_{ir}\rangle  +\langle(\nabla^2 \phi_{ir})\phi_{ir}\rangle
\eea
we obtain a factor inside the $p_1$ integral proportional to $2E_1^2$ and therefore we get  
\bea
-2(m^2 - \lambda^2 \alpha)\langle \phi^2_{ir}\rangle +\langle\phi_{ir} \nabla^2 \phi_{ir}\rangle  +\langle(\nabla^2 \phi_{ir})\phi_{ir}\rangle
& \approx &  -\lambda^2\langle \phi^2_{ir}\rangle \int_{uv}\frac{1}{8}\frac{d^d p_3}{(2\pi)^d}\frac{1}{E^3_3} + 2 \lambda^2 \alpha \langle \phi^2_{ir}\rangle \nonumber\\
& = & \lambda^2\alpha \langle \phi^2_{ir}\rangle \nonumber \\
\label{eq:phidiv}
\eea
as the ${\cal{O}}(\lambda^2)$ contribution at leading order in $\Lambda_0$.  Here we have used $(\ref{phi2})$. 
The term proportional to $\beta$ in (\ref{eq:eomdiv})\ is cancelled by the ${\cal{O}}(\lambda^2)$ term in $\pi_{ir}^2$, reminiscent of the story for the Caldeira-Leggett model. For this calculation we need the contraction:
\bea
 \pi_{ir}(x)\phi_{ir}(z)=i\int \frac{d^d p}{(2\pi)^d}\frac{e^{ip(x-z)}}{2}\left[\Theta(t-z^0)e^{iE_p(t-z^0)}-\Theta(z^0-t)e^{-iE_p(t-z^0)} \right] \nonumber \\
\eea
The relative sign, as compared to the $\phi-\phi$ contraction, is crucial here. The result at ${\cal{O}}(\lambda^2)$ is:
\bea
\langle\pi_{ir}^2\rangle\Big|_{\lambda^2} \!\!&=&\!\!+\frac{\lambda^2}{8}
 \!\!\int_{ir} \frac{d^d p_1}{(2\pi)^d}\int_{uv} \frac{d^d p_3}{(2\pi)^d}\frac{\Theta(\Lambda-|p_3+p_1|)}{2E_3\, 2E_4}\nonumber \\
 &&\int\!\!\int_{-\infty(1-i\epsilon)}^{\infty(1-i\epsilon)}dy^0\,dz^0 g^{-}_1(t-y^0)g^{-}_1(t-z^0)g^+_3(y^0-z^0)g^+_4(y^0-z^0)\nonumber \\
\eea
Performing the integrals over time, we find:
\bea
&&\int_{-\infty(1-i\epsilon)}^{\infty(1-i\epsilon)}dy^0\,dz^0 g^+_1(t-y^0)g^+_2(t-z^0)g^+_3(y^0-z^0)g^+_4(y^0-z^0)\nonumber \\
&=&-\frac{2}{E_1(E_1+E_3+E_4)}+\frac{2}{(E_1+E_3+E_4)^2}\nonumber\\
& \approx & -\frac{2}{E_1(E_3+E_4)}+\frac{4}{(E_3+E_4)^2} \nonumber \\
\eea
The upshot is that for $\pi^2$ we get a subleading extra piece 
\bea
\langle\pi_{ir}^2\rangle&=&\!\! -\frac{\lambda^2}{2}
 \!\!\int_{ir} \frac{d^d p_1}{(2\pi)^d} \frac{1}{2E_1}\int_{uv} \frac{d^d p_3}{(2\pi)^d}\frac{\Theta(\Lambda-|p_3+p_1|)}{2E_3\, 2E_4(E_3+E_4)}\nonumber \\
 && \qquad +\frac{\lambda^2}{2}
 \!\!\int_{ir} \frac{d^d p_1}{(2\pi)^d} \int_{uv} \frac{d^d p_3}{(2\pi)^d}\frac{\Theta(\Lambda-|p_3+p_1|)}{2E_3\, 2E_4(E_3+E_4)^2}
\eea
Therefore, using the same approximations that for $\phi^2$ we get
\bea
2\langle\pi_{ir}^2\rangle \approx -\lambda^2\alpha \langle \phi^2_{ir}\rangle + \lambda^2 \beta \int_{ir} \frac{d^d p}{(2\pi)^d}
\eea
The term proportional to $\alpha$ cancels the divergence in (\ref{eq:phidiv}); the term proportional to $\beta$ is cancelled by the final term in (\ref{eq:eomdiv}).

It appears that, as in \S2, the divergence in the master equation is required to properly account for the renormalization of $\pi^2$.  However, we have been calculating the Schwinger-Dyson equation for vacuum expectation values, whereas the master equation in \S3.2\ was not constructed in the vacuum. We now turn to studying the operator equations of motion in the factorized initial state we considered in that section.  We find that the essential story remains the same, but terms proportional to $\beta$ are spread out over the various components of (\ref{eq:ourcomposite}).

%
%

\subsection{Composite operators in factorized state}

Our previous treatment has made it plausible that the divergence in the master equation serves to ensure that the operator equations of motion close on suitably regularized operators.  However, we are still comparing, if not apples and oranges, oranges and grapefruits: the divergence structure of matrix elements of operators in the vacuum state with the divergence structure of the master equation computed in a factorized initial state.  Since the latter will contain excitations of arbitrarily high energy and momentum with respect to the former, it is possible that this divergence structure could be different: a similar story holds for ``$\alpha$-vacua" in de Sitter space ({\it cf.} \cite{Kaloper:2002cs}\ and references therein).

We will compute the expectation value of the operator
\be
	\vev{{\cal O}} = 2\langle \pi_{ir}^2\rangle-2m^2\langle \phi_{ir}^2\rangle+\langle\phi_{ir} \nabla^2 \phi_{ir} +(\nabla^2 \phi_{ir})\phi_{ir}\rangle\ , \label{eq:uncorrop}
\ee
appearing in (\ref{eq:eomfrommaster}), in the factorized state $\ket{\Psi(t = 0)} = \ket{\psi}_{ir}\ket{0}_{uv}$ used in \S3.2.  We will find that it has a time-independent divergent term which cancels the explicit divergences in (\ref{eq:eomfrommaster}). Interestingly, these divergences are distributed differently across the different terms in ${\cal{O}}$ as compared to the vacuum state, but their sum is the same as in the vacuum state.  We will also find additional transient divergences in ${\cal{O}}$ which are not so cancelled.

%


We work in the interaction picture, and can represent the factorized initial state by studying the pure state density matrix of the full coupled theory (including IR and UV modes):
\be
\sigma_I(t) = T\ e^{-i \lambda \int_0^t d t' V(t')}\sigma(0)\  T^* e^{i \lambda \int_0^{t} dt'' V(t'')}
\ee
Expanding in $\lambda$, we find:
\begin{eqnarray} 
\sigma_I(t) &=& \sigma(0) - i \lambda \int_0^t dt' \[V(t'),\sigma(0)\] \nonumber\\
& & \qquad \qquad -\lambda^2 \int_0^t dt'\int_0^{t'} dt''\ V(t') V(t'') \sigma(0) + \sigma(0) V(t'') V(t')  \nonumber \\
& & \qquad \qquad + \lambda^2 \int_0^t dt' \int_0^t dt'' V(t') \sigma(0) V(t'') \, , \label{eq:pertex}
\end{eqnarray}
to order ${\cal{O}}(\lambda^2)$. The expectation value of any operator will then be given by:
\bea
\langle \mathcal{O}\rangle=\Tr(\mathcal{O}(t)\sigma_I(t))
\eea
The divergent parts appear at ${\cal O}(\lambda^2)$, and since we are interested in a universally divergent piece we will consider our initial state to be the ``factorized" vacuum, therefore
\bea
\vev{{\cal O}}\Big|_{\lambda^2} = -\lambda^2 \int_0^t dt'\int_0^{t'} dt''\ \left(\langle \mathcal{O}(t)V(t') V(t'') \rangle + \langle V(t'') V(t') \mathcal{O}(t)\rangle \right)   \nonumber \\
+ \lambda^2 \int_0^t dt' \int_0^t dt'' \langle  V(t'') \mathcal{O}(t)V(t')\rangle \label{eq:pertopeom2}\,.
\eea
In this equation the only terms in $V$ that will give divergent contributions are the ones that couple the UV and IR oscillators. It should be clear that the divergences will come from terms with UV oscillators in them.  However, bubble diagrams will cancel from (\ref{eq:pertopeom2}), since this state has the proper normalization at ${\cal{O}}(\lambda^2)$.  Tadpole terms coming from $\phi_{uv}^3$ terms in one power of $V$ and $\phi_{ir}^2\phi_{uv}$ in the other, are removed if we take $V_I$ to be normal ordered. As in \S3.2, the only divergent terms will come from the $\phi_{ir}\phi_{uv}^2$ part of $V$, generating intermediates states with two UV particles whose momenta are nearly back-to-back.  

Before continuing, we should note that the computation of the vacuum expectation value in perturbation theory, following the previous section \S3.3, is nearly identical to (\ref{eq:pertopeom2}).  The crucial differences are that in that section we considered time-ordered expectation values, and the time integrals were over a contour in complex time rotated slightly away from the real axis, so as to project onto the vacuum.  Here the expectation values are unordered, and the time integrals are over a finite interval.

%

Let us look at the expectation value of each term in ${\cal{O}}$. First consider $\vev{\phi_{ir}^2}$.  The second term on the right hand side of (\ref{eq:pertopeom2})\ is:
\bea
&&-\lambda^2 \int_0^t dt'\int_0^{t'} dt''\ \langle V(t'') V(t') \phi^2(x,t)\rangle\nonumber \\
&& \qquad =-\frac{\lambda^2}{4}\int_0^t dt'd^dx'\int_0^{t} dt'' d^dx''\langle \phi_{ir}(x'',t'')\phi_{ir}(x',t')\phi^2_{ir}(x,t) \rangle \nonumber\\ 
& & \qquad \qquad \times \langle\phi_{uv}^2(x'',t'') \phi_{uv}^2(x',t') \rangle \nonumber \\
\eea
The UV and IR vacuum correlators are not time ordered, therefore are equal to 
\bea
\langle\phi_{uv}(x'',t'') \phi_{uv}(x',t') \rangle =\int_{uv} \frac{d^dk}{(2\pi)^d2\w_k}e^{ik(x''-x')-i\w_k(t''-t')} 
\eea
and similarly for the ir. Writing all the propagators, doing the time integrals,
and doing the time averaging, we get:
\bea
&&-\lambda^2 \int_0^t dt'\int_0^{t'} dt''\ \langle V(t'') V(t') \phi^2(x,t)\rangle_{div}\Big|_{\lambda^2}\nonumber \\
&&\qquad  \approx \frac{\lambda^2 \alpha }{8}\int_{ir} \frac{d^dp}{(2\pi)^d}\frac{1-e^{2i\w_p t}}{\w_p^3}-\frac{\lambda^2 \beta }{8}\int_{ir} \frac{d^dp}{(2\pi)^d}\frac{1+e^{2i\w_p t}}{\w_p^2}
\eea
Therefore the two first terms on the right hand side of (\ref{eq:pertopeom2})\ in the evaluation of $\langle \phi^2\rangle$ are equal to 
\bea
\frac{\lambda^2 \alpha }{2}\int_{ir} \frac{d^d p}{(2\pi)^d}\frac{\sin^2(\w_p t)}{\w_p^3}-
\frac{\lambda^2 \beta }{2}\int_{ir} \frac{d^d p}{(2\pi)^d}\frac{\cos^2(\w_p t)}{\w_p^2}
\eea
Note that the remaining time dependence is with IR frequencies; thus, they will survive our time averaging.
The remaining term on the right hand side of (\ref{eq:pertopeom2}) is:
\bea
&&\lambda^2 \int_0^t dt'\int_0^{t} dt''\ \langle V(t') \phi^2 (x,t)V(t'') \rangle\nonumber \\
&& \qquad =\frac{\lambda^2}{2}\int_{ir} \frac{d^dp}{(2\pi)^d}\frac{1}{\w_p^2}\int_{uv} \frac{d^dk}{(2\pi)^d}\frac{\Theta(\Lambda-|p+k|)}{2\w_k 2\w_{p+k}(\w_k+\w_{p+k}+\w_p)^2} \nonumber \\
&&\qquad \quad \approx \frac{\lambda^2 \beta}{2}\int_{ir} \frac{d^dp}{(2\pi)^d}\frac{1}{\w_p^2}
\eea
Therefore, we conclude that the leading divergences in $\langle \phi^2\rangle$ at ${\cal{O}}(\lambda^2)$ are:
\bea
\langle \phi^2\rangle_{div}\Big|_{\lambda^2} & \approx & \frac{\lambda^2 \alpha }{2}\int_{ir} \frac{d^d p}{(2\pi)^d}\frac{\sin^2(\w_p t)}{\w_p^3}-\frac{\lambda^2 \beta }{2}\int_{ir} \frac{d^d p}{(2\pi)^d}\frac{\cos^2(\w_p t)}{\w_p^2} \nonumber\\
& & \qquad \qquad + \frac{\lambda^2 \beta}{2}\int_{ir} \frac{d^dp}{(2\pi)^d}\frac{1}{\w_p^2} + \ldots
\eea
In five spatial dimensions there will be a log divergence which we leave for future work.

Similar calculations yield
\bea
\langle \pi^2\rangle_{div}\Big|_{\lambda^2} & \approx &  -\frac{\lambda^2 \alpha }{2}\int_{ir} \frac{d^d p}{(2\pi)^d}\frac{\sin^2(\w_p t)}{\w_p}+\frac{\lambda^2 \beta }{2}\int_{ir} \frac{d^d p}{(2\pi)^d}\cos^2(\w_p t)\nonumber\\
& & \qquad \qquad+\frac{\lambda^2 \beta}{2}\int_{ir} \frac{d^dp}{(2\pi)^d}\ .\nonumber\\
\eea
Note that if we write $\cos^2 \omega_p t = \half + \half \cos 2 \omega_p t$ as the sum of a constant part and a part oscillating about zero, the total $\beta$-dependent constant term willl be different from the vacuum answer.  The remaining oscillating term will contribute a transient that dies off for $t \gg \Lambda^{-1}$ and
\bea
\langle\phi \nabla^2 \phi +(\nabla^2 \phi)\phi\rangle&=&-\lambda^2 \alpha \int_{ir} \frac{d^d p}{(2\pi)^d}\frac{|p|^2\sin^2(\w_p t)}{\w_p^3}\nonumber\\
& & \quad +\lambda^2 \beta \int_{ir} \frac{d^d p}{(2\pi)^d}\frac{|p|^2\cos^2(\w_p t)}{\w_p^2}\nonumber \\
&& \quad -\lambda^2 \beta\int_{ir} \frac{d^dp}{(2\pi)^d}\frac{|p|^2}{\w_p^2}\nonumber \\
\eea
Adding all of these up, we find:
\bea
& & \vev{{\cal{O}}}_{div}\Big|_{\lambda^2} = -2 \lambda^2 \alpha\int_{ir} \frac{d^d p}{(2\pi)^d}\frac{\sin^2(\w_p t)}{\w_p}+2\lambda^2\beta \int_{ir} \frac{d^d p}{(2\pi)^d}\cos^2(\w_p t) \nonumber \\
& & \qquad \qquad = -\lambda^2 \alpha\int_{ir} \frac{d^d p}{(2\pi)^d}\frac{1}{\w_p}+\lambda^2 \alpha\int_{ir} \frac{d^d p}{(2\pi)^d}\frac{\cos(2\w_p t)}{\w_p}\nonumber \\
&& \qquad \qquad \qquad +\lambda^2 \beta \int_{ir} \frac{d^d p}{(2\pi)^d}+\lambda^2\beta \int_{ir} \frac{d^d p}{(2\pi)^d}\cos(2\w_p t)
\eea
Note that the constant parts of this divergence are identical to what we computed in the vacuum -- however, they are distributed differently amongst the different parts of ${\cal{O}}$.
 

The divergent part of the right hand side of (\ref{eq:eomfrommaster}) is thus:
\bea
2\langle \pi^2\rangle-2(m^2-\lambda^2\alpha)\langle \phi^2\rangle+\langle\phi \nabla^2 \phi +(\nabla^2 \phi)\phi\rangle-\lambda \langle \phi^3\rangle-\lambda \beta \int_{ir} \frac{d^d p}{(2\pi)^d}\nonumber \\
=  \lambda^2 \alpha\int_{ir} \frac{d^d p}{(2\pi)^d}\frac{\cos(2\w_p t)}{\w_p}+\lambda^2\beta\int_{ir} \frac{d^d p}{(2\pi)^d}\cos(2\w_p t) + subleading\ldots\nonumber \\
\eea
The explicit time-independent divergence in the Hamiltonian part of the master equation thus renormalizes the mass, while the time-independent divergence in the non-Hamiltonian part ensures that the operator equations renormalize the composite operator ${\cal{O}}$. We are left with transient terms, oscillating about zero, which will decay when $t \gg \Lambda^{-1}$, at which time the integrands of the frequency integrals will be rapidly oscillating. These are artifacts of our singular initial state.

As an additional check that these transients should be present, we explicitly compute the leading divergences in 
$d_t\langle \phi \pi +\pi \phi\rangle$: 
\bea
\frac{d}{dt}\langle \phi \pi +\pi \phi\rangle = \lambda^2 \alpha\int_{ir} \frac{d^d p}{(2\pi)^d}\frac{\cos(2\w_p t)}{\w_p}+\lambda^2\beta\int_{ir} \frac{d^d p}{(2\pi)^d}\cos(2\w_p t)\,.
\eea
these match the remaining transients on the right hand side of (\ref{eq:eomfrommaster}).

\section{Conclusions}

We close with a few comments and further directions for research. First, a full theory of the renormalization group acting on the quantum master equation, upon successive coarse-grainings, should be worked through: some of the divergences should be removed, as here, by renormalizing composite operators and their OPE coefficients.  This should match a Wilsonian-type treatment of the path integral in the Feynman-Vernon approach, along the lines of \cite{Lombardo:1995fg,Avinash:2017asn}.  

Note that in this work, we have chosen our boundary conditions in the UV (that the theory be a closed quantum system, with a Hamiltonian master equation), following Wilsonian renormalization, and successively coarse grained the theory. This is in distinction to standard treatments of renormalization in which the boundary conditions are set at some physical renormalization point (perhaps corresponding to the energy at which actual measurements are made), and one studies the flow of couplings under the change of this renormalization point. This latter approach is the philosophy taken in \cite{Avinash:2017asn}. In this case, it is natural to study flow on the space of non-Hamiltonian master equations (or influence functionals) without any prejudice as to the nature of the high-energy theory.


In computing the master equation, we have stuck to the case that the initial state factorizes between the observed and unobserved parts of the full quantum system.  In this case, the master equation takes the well-known form we have discussed here. An important step would be to construct the master equation for initial states of Born-Oppenheimer type, in which the high-energy components are systematically projected out, extending the work of \cite{2014AnPhy.345..141D,2014arXiv1405.2077D}.

Third, the calculations here are all in perturbation theory, for a weakly coupled theory.  Calculations of the master equation at large $N$, via direct large-N techniques or, where applicable, via holography, would be of great interest.  

\vskip.8cm
	
{\bf Acknowledgments}: 
We would like to particularly thank Vladimir Rosenhaus for early collaboration on this project, and for many useful comments and corrections (any remaining mistakes  and misconceptions are, of course, ours). We would also like to thank Matthew Headrick, Bei-Lok Hu, Ted Jacobson, and R. Loganayagam for useful discussions.  Parts of this work was carried out while A.L. was at the Aspen Center for Physics for the ``Primordial Physics" workshop, which is supported by National Science Foundation grant PHY-1066293; at the NORDITA workshop on ``Black Holes and Emergent Spacetime"; and at the ICTS workshop on ``Classical and Quantum Information", which was supported in part by the NSF IGERT grant ``Geometry and Dynamics", and in part by the NSF grant NSF-OISE-1243369. Part of this work was carried out while A.L. and C.A. attended the ``Entanglement in Strongly-Correlated Quantum Matter" and ``Quantum Gravity Foundations: UV to IR" workshops at the KITP.  They would like to thank the organizers and staff of these workshops for stimulating and productive working environments. A.L. and C.A. are supported in part by DOE grant DE-SC0009987. C.A. is also supported in part by the NSF CAREER awards PHY10-53842, PHY11-25915 and PHY16-20628.

\eject

\addcontentsline{toc}{section}{Appendices}
\section*{Appendices}

\vskip .5cm

\appendix

\section{Exact vacuum two-point correlators in the Caldeira-Leggett model and its factorized counterpart }

In \S2\ we found that while divergences in the Heisenberg equations of motion for $X,P$ are removed by a Born-Oppenheimer-type choice of state, the equations for quadratic operators do contain divergences that appear to be intrinsic to the system.  In particular, the expectation value of $P^2$ can be shown to be divergent, as pointed out for example in \S3.6\ of \cite{breuer2007theory}, in the case of the factorized initial state.  

To make the case that these divergences are not artifacts of the high-energy components of the factorized initial state, in section (\ref{TV})\ we compute $\vev{P(t)P(t')}$ in the true vacuum of the theory, and show that there is an operator product singularity as $t \to t'$, such that the composite operator $P^2$ is divergent and requires renormalization. In 
section (\ref{F-S})\ we then compute the expectation value of $P^2$ in the factorized (zero-temperature) state. In fact, we will find that in both cases we get the same logarithmic divergence (the computations are in different schemes, but the divergences are logarithmic, and the coefficients are identical).

\subsection{Exact vacuum\label{TV}}

In understanding the OPE singularity structure of $P(t) P(t')$, it suffices to compute the Feynman propagator.
Since we are studying a Gaussian Lagrangian, we can compute the vacuum correlation functions of the IR oscillators exactly. 

The direct construction of the vacuum is cumbersome, but for the Feynman propagator, the standard $\epsilon$-prescription in the path integral projects the noninteracting vacuum state $\ket{0}_{IR}\ket{0}_{UV}$ onto the true vacuum.  We thus start with the Lagrangian
\be
	{\cal L} = \half M {\dot X}^2 - \half M (\Omega^2 - i \epsilon) X^2 
	+ \half \sum_i m_i \left({\dot x}_i^2 - (\omega_i - i \epsilon) x_i^2\right) + \sum_i C_i x_i X
\ee
We wish to compute
\be
	Z[J] = {\cal N}\ \int DX \prod_i Dx_i e^{i \int dt \left[ {\cal L} + J(t) X(t)\right]}
\ee
It is straightforward to integrate out $x_i$, to get
\begin{eqnarray}
	& & Z[J] = {\cal N}\ \int DX \exp\left\{ i \int dt \left[ \half M {\dot X}^2 - \half M (\Omega^2 - i \epsilon) +J\,X \right]\right.\nonumber\\
	& & \qquad \qquad \qquad \qquad \left.	+ \frac{i}{2} \sum_k \frac{C_k^2}{m_k} \int dt dt' X(t) O_k(t,t') X(t')\right\}
\end{eqnarray}
where
\be
	O_k(t,t') = \int_{-\infty}^{\infty} \frac{d\omega}{2\pi} \frac{e^{- i \omega(t-t')}}{\omega^2 - \omega_k^2 + i\epsilon}\ .
\ee
Note that we are being cavalier about the normalization factor ${\cal N}$, and absorbing constant multiplicative factors after each integration.

If we write the path integral entirely in terms of the Fourier modes ${\tilde X}(\omega) = \int dt e^{i \omega t}X(t)$, we find
\begin{eqnarray}
	Z[J] & = & {\cal N}\ \int D{\tilde X} \exp\left\{ \frac i2 \int \frac{d\omega}{2\pi} {\tilde X}(\omega){\tilde X}(-\omega)
		\left[ M \omega^2 - M (\Omega^2 - i \epsilon) \right.\right.\nonumber\\
		& & \qquad \qquad \qquad \qquad \qquad - \left. \left. \sum_k \frac{\sigma_k^2}{\omega^2 - \omega_k^2 + i\epsilon} \right]\right\}  \nonumber \\
		&& \qquad \times \exp\{  \frac i2 \int \frac{d\omega}{2\pi} \(\tilde{J}(\w) \tilde{X}(-\w) + \tilde{X}(\w) \tilde{J}(-\w)\) \}   \label{eq:partint}
\end{eqnarray}
where $\sigma_k^2 = C_k^2/m_k$ and $\tilde{J}(\w)$ are the source Fourier modes.

To continue further, we must specify the spectrum of the environment oscillators $x_k$, and perform the sum over $k$.  As in \S2, we consider the  continuous ``Ohmic" spectrum \cite{Caldeira:1982iu}\ down to zero frequency, such that
\be
	\sum_k \sigma_k^2 f(\omega_k) \equiv \frac{4\gamma M}{\pi} \int_0^{\Lambda} d\tom\,  \tom^2 f(\tom)
\ee
which implicitly defines $\gamma$. Here $\Lambda$ is an ultraviolet cutoff. We can put the divergent part of the sum over $k$/integral over $\tom$ in (\ref{eq:partint}):
\be
	- \sum_k \frac{\sigma_k^2}{\omega^2 - \omega_k^2+i\eps } = \frac{4 \gamma M}{\pi} \int_0^{\Lambda}d\tom \left[ 1 + \frac{\omega^2 - i\epsilon}{\tom^2 - \omega^2 - i \eps}\right]\ .\label{eq:massrenint}
\ee
The first term is linearly divergent, and the second term convergent, as $\Lambda \to \infty$. This integral will appear in the denominator of a spectral integral defining $\vev{X(t_2)X(t(1)}$, our object of interest.  If we wish to ignore terms in that two-point function which vanish as $\Lambda \to \infty$, we can extend the integral out to infinity in the second term.
Since this is even, we can further extend the integral over the real line, and close in either the upper or lower half plane, to find:
\be
	- \sum_k \frac{\sigma_k^2}{\omega^2 - \omega_k^2} = \frac{4 \gamma M \Lambda}{\pi} +2 i M \gamma |\omega|
\ee
The first term leads to a renormalization of $\Omega$.  
We define $\Gamma = \frac{\pi \gamma}{2 M}$, and $\Omega_r^2 = \Omega^2 - \frac{4 \gamma M}{\pi} \Lambda$. We will consider theories for which $\Omega_r^2 > 0$. The resulting expression for the path integral is
\be
Z[J] = {\cal N}\ \exp \left\{ \frac{1}{2} \int_{-\infty}^{\infty} \frac{d\omega}{2\pi} {\tilde J}(\omega)\tilde{G}(\w){\tilde J}(-\omega)\right\}\label{eq:amplitude}
\ee
where 
\bea
\tilde{G}(\w)=\frac{i}{M (\omega^2 - \Omega_r^2 + i \epsilon + 2 i \gamma |\omega|)}\ ,
\eea
which we arrive at by completing the square in (\ref{eq:partint})\ and integrating over $\tilde{X}$.
%

The resulting time-ordered vacuum two-point function is, for $t_2 > t_1$, 
\begin{eqnarray}\label{eq:tpfint}
	\bra{0} T\left[ X(t_2) X(t_1)\right] \ket{0} & = & \frac{i}{M} \int_{-\infty}^{\infty} \frac{d\omega}{2\pi} \frac{e^{- i \omega (t_2 - t_1)}}{\omega^2 - \Omega_r^2 + i (2 \gamma |\omega| + \eps)}\nonumber\\
	& = & \frac{i}{2\pi M} \left[ \int_0^{\infty} d\omega \frac{e^{-i \omega(t_2-t_1)}}{\omega^2 - \Omega_r^2 + i (2 \gamma \omega + \eps)}\right.\nonumber\\
	& & \ \ \ \ \ \ \ \ \left.
		+ \int_{-\infty}^0 d\omega \frac{e^{- i \omega(t_2 - t_1)}}{\omega^2 - \Omega_r^2 - i (2 \gamma \omega - \eps)} \right]\nonumber\\
		& \equiv & F_+ + F_-
\end{eqnarray}
At this point we will set $\eps \to 0$: for $\Gamma \neq 0$ the location of the poles of the integrand will be well off the real axis. We compute each of $F_{\pm}$ in turn.

\begin{itemize}

\item $F_+$.  The integrand has a pole  at:
\be
	\omega_{\pm} = - i \gamma \pm \sqrt{\Omega_r^2 - \gamma^2} \equiv - i \gamma \pm Q\ .
\ee
Since $\Delta t = t_2 - t_1 > 0$, we can rotate the integration contour into the negative imaginary axis, picking up the residue of the pole at $\omega_+$:
\be
	F_+ = \frac{e^{- i \omega_+ \Delta t}}{2 M Q} - \frac{1}{2\pi M} \int_0^{\infty} dw \frac{e^{-w \Delta t}}{w^2 + \Omega_r^2 - 2 \gamma w}\ .
\ee

\item $F_-$.  The contour can be rotated to the negative imaginary axis, without passing any poles:
\be
	F_- = \int_0^{\infty} dw \frac{e^{- w \Delta t}}{w^2 + \Omega_r^2 + 2 \gamma w}
\ee

\end{itemize}

Combining $F_{\pm}$ and performing a bit more algebra, we find
\begin{eqnarray}
	\vev{X(t_2) X(t_1)} & = & \frac{e^{- z_{++} \Delta t}}{2 M Q} + \frac{1}{4\pi i M Q}\int_0^{\infty} dw e^{- w \Delta t}\left\{- \frac{1}{w - z_{++}} \right. \nonumber\\
	& & \qquad \qquad \qquad \left.+ \frac{1}{w - z_{+-}}
		+ \frac{1}{w - z_{-+}} - \frac{1}{w - z_{--}}\right\}\nonumber\\
	& = &  \frac{e^{- z_{++} \Delta t}}{2 M Q} + \frac{1}{4\pi i M Q}\left[ - e^{- z_{++} \Delta t} E_1(- z_{++} \Delta t)
	\right.\nonumber\\
	& & \qquad \qquad \qquad + e^{- z_{+-} \Delta t} E_1(- z_{+-}  \Delta t) \nonumber\\
	& & \qquad \qquad \qquad  + e^{ - z_{-+} \Delta t} E_1(-z_{-+}\Delta t) \nonumber\\
	& & \qquad \qquad \qquad \left. - e^{-z_{--}\Delta t} E_1(-z_{--}\Delta t)\right]\nonumber \\ \label{eq:exactxx}
\end{eqnarray}
where
\be
	z_{\sigma\sigma'} = \sigma\gamma + \sigma' i Q\ .
\ee
and we use $E_1$ as defined in \cite{abramowitz1965handbook}.  $E_1$ has a logarithmic branch cut along the negative real axis; we take all of the arguments in (\ref{eq:exactxx}) to lie on the first sheet.

For small arguments, $E_1(z) \sim - \gamma_E - \ln z + z - \frac{z^2}{4} + \ldots$, where $\gamma_E$ is the Euler-Mascheroni constant.  Using this, we find that
\be
	\lim_{\Delta t \to 0} \vev{X(t_2) X(t_1)} = \frac{1}{2 M Q} + \frac{1}{\pi M Q} \tan^{-1}\left(\frac{\gamma}{Q}\right)
\ee
As $\gamma \to 0$ (decoupling $X$ from $x_i$), $\Gamma \to 0$ and $Q \to \Omega$, and this becomes the known width of the ground state of the simple harmonic oscillator.

Since $P(t) = M {\dot X}(t)$, 
\be
	\bra{0} T[P(t_2) P(t_1)] \ket{0} = M^2 \p_{t_2}\p_{t_1} \vev{X(t_2)X(t_1)}\ .
\ee
One can start from (\ref{eq:exactxx}) in integral form and take derivatives with respect to $\Delta t, -\Delta t$ respectively, which leads to 
\bea
-\partial_{\Delta t} \partial_{\Delta t} \vev{X(t_2) X(t_1)}&=&-\frac{z_{++}^2\, e^{- z_{++} \Delta t}}{2MQ} -\frac{1}{4\pi i M Q}\int_0^{\infty} dw e^{- w \Delta t}\left\{- \frac{w^2}{w - z_{++}} \right. \nonumber\\
	& & \qquad \qquad \qquad \left.+ \frac{w^2}{w - z_{+-}}
		+ \frac{w^2}{w - z_{-+}} - \frac{w^2}{w - z_{--}}\right\}\nonumber \\
&&
\eea
The integral term looks divergent: however let us rewrite a single one of the terms in brackets, namely  
\bea
\frac{w^2}{w - z_{++}}=w+z_{++} +\frac{z^2_{++}}{w - z_{++}}
\eea
The term proportional to $w$ cancels trivially between the four terms inside the brackets, as does the term proportional to $z_{++}$, since $-z_{++}+z_{+-}+z_{-+}-z_{--} =0$. Thus, the following replacement is valid inside the brackets:
\bea
\frac{w^2}{w - z_{\sigma \sigma'}}\to \frac{z^2_{\sigma \sigma'}}{w - z_{\sigma \sigma'}}
\eea
which leads to 
\bea
\vev{P(t_2)P(t_1)} &=& -\frac{M z_{++}^2 e^{- z_{++} \Delta t}}{2 Q} + \frac{1}{4\pi i M Q}\left[ - z^2_{++} e^{- z_{++} \Delta t} E_1(- z_{++} \Delta t)
	\right.\nonumber\\
	& & \qquad \qquad \qquad + z^2_{+-} e^{- z_{+-} \Delta t} E_1(- z_{+-}  \Delta t) \nonumber\\
	& & \qquad \qquad \qquad  + z^2_{-+} e^{ - z_{-+} \Delta t} E_1(-z_{-+}\Delta t) \nonumber\\
	& & \qquad \qquad \qquad \left. - z^2_{--} e^{-z_{--}\Delta t} E_1(-z_{--}\Delta t)\right]\nonumber \\
	\eea
Using the expansion of the $E_1$ one find a logarithmic divergence when $\Delta t \to 0$
%
\be
	\vev{P(t_2)P(t_1)}_{t_2 \to t_1} \sim \frac{M Q}{2} \left(1 + \frac{4\gamma}{\pi Q}(1 - \gamma_E) + \frac{2}{\pi} \tan^{-1} \frac{\gamma}{Q}\right) - \frac{2 \gamma M}{\pi} \ln \Omega_r \Delta t
\ee
In the limit $\Gamma \to 0$, $Q \to \Omega$, the first constant term becomes the width in momentum space of the ground state wavefunction of the simple harmonic oscillator.

We should note $Q$ is a function of the renormalized frequency; if we expand $Q$ in small $\gamma$ about the bare frequency $\Omega$, then we will see a linear divergence in $\vev{P^2}$, in addition to the logarithmic one. 

%

\subsection{Factorized state \label{F-S}}

In this section we compute the expectation value of $P^2$ in the factorized initial state $\rho_{UV}(0)\times \rho_{IR}(0)$, where for simplicity we will consider $\rho_{UV}(0)=|0 \rangle \langle 0|$. This gives rise to initial jolt singularities in the Heisenberg equations for $X,P$ and the quadratic operators.  As we argued in \S2, these initial jolts can be projected out by projecting out high-energy components, following the Born-Oppenheimer approximation.  Nonetheless, additional divergences in the equations of motion appear to remain. The presence of OPE divergences in the exact vacuum, found above, gives credence to this view, and indicates that they should be related to operator renormalization.  Thus, we compute the divergence in $P^2(t)$ for large $t$; this result will be used in the main text to show that it will be cancelled exactly by the divergence in the operator equations of motion. In this case the divergence is identical to the vacuum divergence: in general, for states containing substantial high energy excitations, there is no reason for the OPE singularities to match the vacuum OPEs (see \cite{Kaloper:2002cs}\ and references therein).

Taking $t>0$ in (\ref{eq:fullheisx}), we have:
\bea \label{P}
P(t)&=&-e^{-\gamma t}M X_0(\gamma \cos\W_r t+\W_r\sin\W_r t)+e^{-\gamma t}P_0 \left(\cos\W_r t -\gamma \frac{\sin\W_r t}{\W_r}\right)\nonumber \\
&& \qquad +\sum_j\frac{C_j}{\W_r}\int_0^t
ds (\W_r \cos\W_r(t-s)-\gamma\sin\W_r(t-s))e^{-\gamma (t-s)}x_j^{0}(s) \,. \nonumber \\
\eea
Note that the first and second lines depends on the initial conditions for the IR  and UV operators respectively. Since we are interested in the divergent terms in $\langle P^2(t) \rangle$, we can safely ignore the square of the terms depending on the IR operators. On the other hand, the crossing terms will die in the expectation value since they are linear in the creation and annihalation UV operators: that is, $\langle x_j^0(t)\rangle=0$ in our particular initial state.

The upshot is that the potentially divergent terms in $\vev{P^2}$ are:
\bea
\langle P^2(t)\rangle  & \sim & \int_0^t d \tau \int_0^t d\tau' (\cos\W_r \tau-\frac{\gamma}{\W_r}\sin\W_r\tau)(\cos\W_r \tau-\frac{\gamma}{\W_r}\sin\W_r\tau)\,e^{-\gamma \tau} e^{-\gamma \tau'} \nonumber \\
&& \,\frac 12\sum_j C_iC_j \langle \{ x^{0}_i(t-\tau), x^{0}_j(t-\tau')\} \rangle\ . \nonumber \\
\eea
Using 
\bea
\langle x^{0}_l(s)x^{0}_j(t) +x^{0}_j(t)x^{0}_l(s)   \rangle =\delta_{jl}\frac{\cos\w_j(t-s)}{m_j\w_j}\ ,
\eea
we obtain
\bea
\langle P^2(t)\rangle  & \sim & \sum_j \frac{C^2_i}{2m_i \w_i}\Bigg[\left(\int_0^t d \tau  (\cos\W_r \tau-\frac{\gamma}{\W_r}\sin\W_r\tau)\,e^{-\gamma \tau}  \cos\w_i\tau\right)^2\nonumber \\
&& +\left(\int_0^t d \tau  (\cos\W_r \tau-\frac{\gamma}{\W_r}\sin\W_r\tau)\,e^{-\gamma \tau} \sin\w_i\tau \right)^2\Bigg] \ .\nonumber \\
\eea
Since we are not interested here in the finite $t$ effects we can take $t \to \infty$ and use the Drude regularization $F(\w)=1/(1+(\w/\Lambda)^2)$ which leads to
\bea
\langle P^2(t)\rangle  & \sim & -\frac{2\gamma M}{\pi}\log\left(\sqrt{\W_r^2+\gamma^2}/\Lambda \right)+\frac M\pi \left(\frac{\W_r^2-\gamma^2}{2 \W_r}\right)\arctan\left(\frac{\W_r}{\gamma}\right) \nonumber \\
\eea
(We expect the log divergence to be independent of the details of the cutoff). This divergence matches the vacuum OPE singularity $\langle P(t)P(t')\rangle$ exactly.

As above, the second term, when expanded in small $\gamma$ about $\Omega$, contains a linear divergence. The full expression is clearly a function of the renormalized frequency.

\section{Calculations for Born-Oppenheimer initial states \label{B}}

In this appendix we compute the final terms in (\ref{eq:dtxp},\ref{eq:dtpp}) for the initial state
%
%
\be
	\ket{\Psi} = \int dX \Psi(X) \ket{X}_{IR} \ket{0;X}_{UV}\,.
\ee
in which the environment oscillators have been placed in their instantaneous ground state for fixed IR oscillator $X$, following the Born-Oppenheimer approximation. This is the leading order expression in a systematic expansion in which one projects out high-frequency components of $\Psi$ , for the high-frequency oscillators with $\omega_i \gg \Omega$. That said, as discussed in \S2.5, we will choose this initial state even for $\omega_i \lesssim \Omega$, for simplicity and comparison with the other calculations in \S2.   

Taking the final terms in (\ref{eq:dtxp},\ref{eq:dtpp})\ as source terms for the differential equations in $O_k$, we evaluate them by substituting in eq. (\ref{eq:fullheisx}), and its derivative $P(t) = M {\dot X}(t)$.
%
In order to evaluate the last terms in (\ref{eq:dtxp},\ref{eq:dtpp}) we thus need the following quantities:
\bea
\sum_j C_j \vev{{ \cal O} x_j^0(t) + x_j^0(t){ \cal O}}
\eea
with $ \cal O$ $=\{X_S, P_S, \sum_i C_i x_i^0(s) \}$.\\

For ${\cal O} = X_S$, we find:
\bea
& & \sum_j C_j \vev{X_S x_j^0(t) + x_j^0(t) X_S}\nonumber\\
& & \qquad \qquad = 2\sum_j C_j \vev{X_S x_j^0(t)} \nonumber \\
& & \qquad \qquad = 2\sum_j C_j \!\!\int \!\! dX \Psi(X) \Psi^{*}(X) X\langle 0,X|x_j^0(t)|0,X \rangle \nonumber \\
& & \qquad \qquad = 2\sum_j \frac{C^2_j}{m_j \w_j^2} \vev{X_S^2}\cos\w_j t =8\gamma M \delta{(t)}\vev{X^2} \nonumber \\
\eea
where we have used the fact that $\langle 0,X|x_{j,S}^0|0,X \rangle = C_jX/m_j\w_j^2$.\\

For ${\cal O} = P_S$ there is an additional complication: when $P_S$ is expressed as a derivative acting on wavefunctions, there will also be derivative terms acting on $\ket{0;X}$.  Let us consider the first term in  $\sum_j C_j \vev{P_S x_j^0(t) + x_j^0(t) P_S}$:
\bea 
\sum_j C_j \vev{P_S x_j^0(t) }=\sum_j C_j \int dXdY \Psi^{*}(Y) \Psi(X) \langle Y |P_S|X\rangle\langle 0,Y|x_j^0(t)|0,X\rangle \ . \nonumber \\
\eea
Using $\langle Y |P_S|X\rangle=-i\partial_{Y} \delta(Y-X)$, and integrating by parts, we find: 
\bea \label{Px}
\sum_j C_j \vev{P_S x_j^0(t) }=i\sum_j C_j \int dX \Psi(X)\partial_Y(  \Psi^{*}(Y) \langle 0,Y|x_j^0(t)|0,X\rangle )\large|_{Y=X}\,.\nonumber \\
\eea
The term in this equation coming from $\p_Y$ acting on $\Psi^*(Y)$ gives:
\bea \label{primero}
i\sum_j C_j \int dX \Psi(X)\partial_X \Psi^{*}(X)\!\!\!\!\!\!\!\!\! &&\!\!\!\! \langle 0,X|x_j^0(t)|0,X\rangle\, \nonumber \\
&=&\sum_j \frac{C^2_j}{m_j \w_j^2}\cos\w_j t\int dX \Psi(X) X i\partial_X \Psi^{*}(X) \nonumber \\
&=& 4\gamma M \delta(t) \langle  P_S X_S \rangle=4\gamma M\delta(t) \langle P X \rangle\ .
\eea
For the term in (\ref{Px})\ coming from $\p_Y$ acting on $\bra{0;Y}$:
\bea
\label{derivative}
i\sum_j C_j \int dX \Psi(X)\Psi^{*}(X)\!\!\!\!\!\!&&\!\!\!\! \partial_Y \langle 0,Y|x_j^0(t)|0,X\rangle\,\large|_{Y=X} \,. 
\eea
we must compute $\partial_Y \langle 0,Y|x_j^0(t)|0,X\rangle$. Using 
\bea
\langle 0,Y|x_j^0(t)|0,X\rangle=\langle 0,Y|x_{j,S}^0|0,X\rangle\cos\w_j t +\langle 0,Y|p_{j,S}^0|0,X\rangle \frac{\sin\w_j t}{m_j\w_j}\,
\eea
we introduce resolutions of the identity for the UV oscillators in the position basis:
\bea
\label{YxX}
&&\langle 0,Y|x_j^0(t)|0,X\rangle =\Bigg (\int dx_j x_j \langle 0,Y|x_j\rangle \langle x_j|0,X\rangle\cos\w_j t \nonumber \\ &&\qquad \quad -\int dx_j \langle 0,Y|x_j\rangle i\partial_{x_j} \langle x_j |0,X\rangle \frac{\sin\w_j t}{m_j\w_j}\Bigg)\, \nonumber\\
&& \qquad \qquad \times \prod_{k\neq j}\int dx_k\langle 0,Y|\prod_{k\neq j}x_k\rangle \langle \prod_{k\neq j}x_k|0,X\rangle \nonumber \\
\eea
Now the instantaneous ground state wavefunctions for the UV oscillators are: 
\bea
\langle \prod_{k}x_k|0,X\rangle=\prod_k \left( \frac{m_k\w_k}{\pi}\right)^{1/4}\exp\left \{ -\frac12 m_k \w_k\left(x_k-\frac{C_k X}{m_k \w_k^2}\right)^2\right\}
\eea
From this one can see that the contribution of $\partial_Y \langle 0,Y|x_j^0(t)|0,X\rangle$ from the derivative acting on the last line of (\ref{YxX}) is zero, since the resulting Gaussian integral is odd in its argument. For the surviving terms, this last line is normalized to one as $Y \to X$, so that: 
\bea
\label{YX}
\partial_{Y}\langle 0,Y|x_j^0(t)|0,X\rangle \large|_{Y=X} &=&\int dx_j x_j ( \partial_X \langle 0,X|x_j\rangle ) \langle x_j|0,X\rangle\cos\w_j t \nonumber \\
&& -i \int dx_j  ( \partial_X \langle 0,X|x_j\rangle ) \partial_{x_j} \langle x_j |0,X\rangle \frac{\sin\w_j t}{m_j\w_j} \nonumber \\
&=&\frac{C_j}{\w_j}\int dx_j (x_j-C_jX/m_j\w_j^2)^2|\langle x_j|0,X\rangle|^2\nonumber\\
& & \qquad \qquad \times (\cos\w_j t +i\sin\w_j)\nonumber \\
&=&\frac{C_j}{2m_j\w_j^2}(\cos\w_j t +i\sin\w_j t)
\eea
which is independent of $X$. Therefore equation (\ref{derivative}) becomes
\bea
\label{second}
\sum_j \frac{C_j^2}{2m_j\w_j^2}(i \cos\w_j t -\sin\w_j t)=2\gamma M i \delta(t)-2\gamma M \int_0^\infty \frac{d\w}{\pi} \sin \w t \,.
\eea
Equation (\ref{Px}), is obtained by adding (\ref{primero}) and (\ref{second}), that is
\bea
\sum_j C_j \vev{P_S x_j^0(t)}=4\gamma M \delta(t) \vev{PX}+2\gamma M i \delta(t)-2\gamma M\int_0^\infty \frac{d\w}{\pi} \sin \w t
\eea
and therefore
\bea
\sum_j C_j \vev{P_S x_j^0(t)+x_j^0(t) P_S}=4\gamma M \delta(t) \vev{PX+XP}-4\gamma M \int_0^\infty \frac{d\w}{\pi} \sin \w t\,. \nonumber \\
\eea
Finally, we want to evaluate $\sum_{ij} C_i C_j \vev{x_i^0(s)x_j^0(t)+x_j^0(t) x_i^0(s)}$. This is 
\bea
\sum_{ij} C_i C_j \vev{\{x_i^0(s),x_j^0(t)\}}=\int |\Psi(X)|^2\langle 0,X| \{ x_i^0(s),x_j^0(t) \}  |0,X\rangle
\eea
Now if, for fixed $X$, we set 
\be
	x_{j,S}^{0} \equiv y_{j,S}^{0} + \frac{C_j X}{m_j \omega_j^2}
\ee
then $\vev{y_{j,S}^0}=\vev{p_{j,S}^0}=0$, and
\bea
& & \sum_{ij} C_i C_j \vev{\{x_i^0(s),x_j^0(t)\}}\nonumber\\
& & \qquad \qquad =2\sum_{i} \frac{C_i^2}{m_i\w_i^2} \cos \w_i s\sum_j \frac{C_j^2}{m_j\w_j^2}\cos\w_j t\int |\Psi(X)|^2 X^2dX \nonumber \\
&&  \qquad \qquad \qquad \qquad +\sum_{ij} C_i C_j \int |\Psi(X)|^2\langle 0,X| \{ y_i^0(s), y_j^0(t) \}  |0,X\rangle dX \nonumber \\
&& \qquad \qquad \to 32\gamma^2 M^2 \delta_{\Lambda}(t)\delta_{\Lambda}(s) \vev{X^2} +4\gamma M \int_0^{\Lambda} \frac{d\w}{\pi} \w \cos \w (t-s) \nonumber \\ \label{eq:tpker}
\eea
The final line is the expression for the Ohmic spectrum with a hard frequency cutoff.  Here 
\be
	\delta_{\Lambda}(t) = \int_0^{\Lambda} \frac{d\omega}{\pi} \cos(\omega t)
\ee
so that $\lim_{\Lambda\to\infty} \delta_{\Lambda}(t) = \delta(t)$.  We have kept the cutoff explicit for the following reason.  Eq. (\ref{eq:tpker}) appears integrated over $s \in [0,t]$ in (\ref{eq:dtxp},\ref{eq:dtpp}): the integral comes from inserting the exact solution for $P$ derived from (\ref{eq:fullheisx}).  If we take $\Lambda\to\infty$ before doing the integral, the integral over $s$ would pin the integrand at $s = 0$, leaving an explicit term proportional to $\delta(t)$.  However, that delta function also collapses the limits of integration; so we must be more careful in interpreting this integral.

The first term in the last line of (\ref{eq:tpker}) appears in (\ref{eq:dtpp}) via the integral
\be
	32 \gamma^2 \frac{M}{\tOM} \delta_{\Lambda}(t) \int_0^t e^{-\gamma(t-s)} \sin(\tOM(t-s)) \delta_{\Lambda}(s)
\ee
Inserting the integral definition of $\delta_{\Lambda}(s)$, the integrand is a sum of exponentials of $s$ and $t$.  We can do the $s$ integral explicitly.   The result is a set of terms which is finite as $\Lambda \to \infty$, and which vanish at $t = 0$.  If we then take $\Lambda\to\infty$ and replace $\delta_{\Lambda}(t) \to \delta(t)$, this term vanishes.

Returning to (\ref{eq:dtxp}) and (\ref{eq:dtpp}), we first evaluate the latter:
\bea
\sum_jC_j\vev{\{P(t),x_j^0(t)\}}=-8\gamma^2 M^2 \vev{X^2}\delta(t)+8\gamma^2 M^2 \vev{X^2}(3-4\theta(t))\delta(t)\nonumber \\
 +4\gamma M \delta(t)\vev{\{P,X\}}-4\gamma M e^{-\gamma t}(\cos\tilde{\W} t-\frac{\gamma}{\tilde{\W}}\sin\tilde{\W} t)G(t) \nonumber  \\
 +4\gamma M \int_0^t d\tau e^{-\gamma \tau}(\cos\tilde{\W} \tau-\frac{\gamma}{\tilde{\W}}\sin\tilde{\W} \tau)G'(\tau)\nonumber \\ \label{eq:ppsource}
\eea
where 
\bea
G(t)=\int_0^\infty \frac{d\w}
{\pi} \sin\w t
\eea
Equation (\ref{eq:ppsource}) contains the exact jolt term proportional to $\vev{\{P,X\}}$, required to eliminate the jolt in (\ref{eq:dtpp}). The extra ${\cal O}(\gamma^2)$ jolt terms also cancel with each other if we remember that we have adopted the convention $\theta(0) = \half$.  The above expression thus reduces to 
\bea
\sum_jC_j\vev{\{P(t),x_j^0(t)\}} & = & 4\gamma M \delta(t)\vev{\{P,X\}}\nonumber\\
& & \qquad   -4\gamma M e^{-\gamma t}(\cos\tilde{\W} t-\frac{\gamma}{\tilde{\W}}\sin\tilde{\W} t)G(t) \nonumber  \\
& & \qquad  +4\gamma M \int_0^t d\tau e^{-\gamma \tau}(\cos\tilde{\W} \tau-\frac{\gamma}{\tilde{\W}}\sin\tilde{\W} \tau)G'(\tau)\ . \nonumber \\
\eea
The last two terms can be combined via integration by parts and leads to 
\bea
\sum_jC_j\vev{\{P(t),x_j^0(t)\}} & = &
4\gamma M \delta(t)\vev{\{P,X\}}\nonumber\\
& &  +4\gamma M \int_0^t d\tau \frac{d}{d\tau}\left[ e^{-\gamma \tau}(\cos\tilde{\W} \tau-\frac{\gamma}{\tilde{\W}}\sin\tilde{\W} \tau)\right ]G(\tau)\,.\nonumber \\
\eea
The same steps can be followed for the other term, which leads to 
\bea
\sum_jC_j\vev{\{X(t),x_j^0(t)\}}=
8\gamma M \delta(t)\vev{X^2}+\frac{ 4\gamma}{\tilde{\W}}\int_0^t d\tau \frac{d}{d\tau}\left[ e^{-\gamma \tau}\sin\tilde{\W} \tau \right ]G(\tau)\,.\nonumber \\
\eea
and with these the master equations  (\ref{eq:dtxp},\ref{eq:dtpp}) become
\begin{eqnarray}
\label{f1}
	\frac{d}{dt} \vev{O_2} & = & \frac{2}{M} \vev{O_3} - 2 M \Omega_r^2 \vev{O_1} - 2\gamma \vev{O_2} \nonumber\\
	& & \qquad +\frac{ 4\gamma}{\tilde{\W}}\int_0^t d\tau \frac{d}{d\tau}\left[ e^{-\gamma \tau}\sin\tilde{\W} \tau \right ]G(\tau) \nonumber \\ 
\end{eqnarray}
and
\bea
\label{f2}
	\frac{d}{dt} \vev{O_3} & = & - M \Omega_r^2\vev{O_2} - 4\gamma M \vev{O_3} \nonumber\\
	& & \qquad + 4\gamma M \int_0^t d\tau \frac{d}{d\tau}\left[ e^{-\gamma \tau}(\cos\tilde{\W} \tau-\frac{\gamma}{\tilde{\W}}\sin\tilde{\W} \tau)\right ]G(\tau)\,.\nonumber\\
\eea

\section{Linearly coupled oscillators at high temperature \label{LargeT}}

As a check on our calculations, we would like to reproduce the result of the original Caldeira and Leggett model, with the environment placed at high temperature, from our perturbative treatment. This is not a particularly new calculation -- see for example the extensive discussion of this model in \cite{breuer2007theory}, but we wish to highlight the relationship between that well-studied limit and our own zero-temperature results.

In that work, one considers the initial density matrix to be factorized between $X$ and $x_i$, with the oscillators $x_i$ described by a thermal density matrix
\be
	\rho = \frac{1}{Z(\beta)} e^{-\beta \sum_i H_i}
\ee
where $H_i = \sum_i \left(\frac{p_i^2}{2m_i} + \half m_i \omega_i^2 x_i^2\right)$, the spectrum of $x_i$ is taken to be the Ohmic spectrum with a cutoff frequency $\L$, and $\frac{1}{\beta} = T \gg \L$.  This is an unusual limit from a Wilsonian point of view, but succeeds in reproducing a Langevin-like equation for $\vev{X}$ that describes Brownian motion.

In this limit the UV vacuum green functions $g_j(t-t')=e^{-i\w_j(t-t')}/2m_j\w_j$ are replaced by the thermal green functions 
\bea
g_j(t-t')=\frac 1{2m_j \w_j}(\coth(\beta \w_j/2)\cos\w_j(t-t')-i\sin\w_j(t-t'))
\eea
In the large-$T$ limit this becomes
\bea
g_j(t-t') \approx \frac{1}{2m_j \w_j}\(\frac{2 T}{\w_j}\cos\w_j(t-t')-i\sin\w_j(t-t')\)\,,
\eea

This modification shifts the real part of the function $\int d\tau F(\tau)X(-\tau)$ that appears in equation (\ref{Ftt'2}), such that its right hand side is replaced by\footnote{The function $\tilde{F}(\tau)$ is defined similarly but with the following change $e^{-i\w\tau}\to 2T/w \cos (\w \tau)-i\sin(\w\tau)$}
\bea
\label{Ftt'3}
\int_0^\infty d\tau \tilde{F}(\tau)X(-\tau )&=&\left(kT -i\frac\L2 \right)X-\left(\frac{T}{M\Lambda}-\frac{i}{2M}\right) P
\eea
and we obtain the Caldeira-Leggett result plus an extra term in the master equation
\bea
\frac{d}{dt}\rho(t)&=&\frac{1}{i}\left[H_{ren},\rho(t) \right]+\frac{\gamma}{2i}[\{X,P\},\rho(t)]-2\gamma M T [X,[X,\rho(t)]] \nonumber \\
&& +\frac{\gamma}{i} ([X,\rho(t)P]-[P,\rho(t)X]) +\frac{\gamma T }{\Lambda}([X,[P,\rho(t)]]+[P,[X,\rho(t)]])\,. \nonumber \\
\eea
The final cutoff-dependent term does not vanish in the $\Lambda\to\infty$ limit, as we are taking $T \gg \Lambda$.  Instead, this term can be argued to be smaller than the others (see {\it e.g.}\ \cite{breuer2007theory}): if $P \sim M \Omega X$, this term is small relative to the term $-2\gamma M T [x,[x,\rho]]$ by a factor of $\Omega/\Lambda$.  In the meantime, the logarithmic divergences we found in the zero-temperature limit are sub-leading by a factor $\frac{\Lambda\ln \Lambda}{T}$ in this high-temperature limit.

\bibliographystyle{utphys}
\bibliography{divergences_refs}

\end{document}